\documentclass[onecolumn,authoryear]{els-mrw}

\usepackage{amsmath,amssymb,amsfonts,amsthm,makeidx,graphicx}
\usepackage{txfonts}
\usepackage{helvet}
\usepackage{xcolor}
\usepackage[colorlinks=true,citecolor=blue, urlcolor=blue, ]{hyperref}

\newcommand{\tzo}{T\.ZO}
\newcommand{\msol}{M$_{\odot}$}
\newcommand{\rsol}{R$_{\odot}$}
\newcommand{\lsol}{L$_{\odot}$}
\newcommand{\teff}{T$_{\mathrm{eff}}$}

\def\aj{AJ}
\def\apj{ApJ}
\def\apss{Astrophys.~Space Sci.}
\def\apjl{ApJL}
\def\apjs{ApJS}

\def\aap{A\&A}
\def\aapr{A\&A~Rev.}
\def\araa{ARAA}
\def\mnras{MNRAS}

\def\planss{Planet. Space~Sci.}

\def\prc{PRC}

\def\actaa{Act.~Astr.}
\def\nat{Nature}

\def\sovast{Sov.~Ast.}

\begin{document}

\chapter{Thorne-\.Zytkow Objects}\label{chap1}

\author[1]{Anna O'Grady}%
\author[2,3,4]{Takashi J. Moriya}%
\author[5]{Mathieu Renzo}%
\author[6]{Alejandro Vigna-G\'omez}%

\address[1]{\orgname{Carnegie Mellon University}, \orgdiv{McWilliams Center for Cosmology, Department of Physics}, \orgaddress{5000 Forbes Avenue, Pittsburgh, PA 15213, USA}}
\address[2]{\orgname{National Astronomical Observatory of Japan}, \orgaddress{2-21-1 Osawa, Mitaka, Tokyo 181-8588, Japan}}
\address[3]{\orgname{Graduate Institute for Advanced Studies, SOKENDAI}, \orgaddress{2-21-1 Osawa, Mitaka, Tokyo 181-8588, Japan}}
\address[4]{\orgname{Monash University}, \orgdiv{School of Physics and Astronomy}, \orgaddress{Clayton, Victoria 3800, Australia}}
\address[5]{\orgname{University of Arizona}, \orgdiv{Department of Astronomy and Steward Observatory}, \orgaddress{933 N. Cherry Avenue, Tucson, AZ 85721, USA}}
\address[6]{\orgname{Max-Planck-Institut f\"ur Astrophysik}, \orgaddress{Karl-Schwarzschild-Str.~1, 85748 Garching, Germany}}

\articletag{Chapter Article tagline: update of previous edition,, reprint..}

\maketitle

\begin{glossary}[Glossary]
\term{Common Envelope Evolution} A phase of binary evolution where two stars share the same envelope\\
\term{Metallicity} The mass fraction of elements heavier than H and He\\
\term{Supernova} The luminous explosion of a massive star\\
\term{Thorne-\.Zytkow Object} The product of a merger of a neutron star with the core of a red supergiant\\
\end{glossary}

\begin{glossary}[Nomenclature]
\begin{tabular}{@{}lp{34pc}@{}}
NS & Neutron star\\
RSG & Red supergiant\\
T\.ZO & Thorne-\.Zytkow Object\\
Z & Metallicity\\
\end{tabular}
\end{glossary}

\begin{abstract}[Abstract]
Interacting binary star systems play a critical role in many areas of astrophysics. One interesting example of a binary merger product are Thorne-\.Zytkow Objects (T\.ZOs), stars that look like red supergiants but contain neutron stars at their cores. T\.ZOs were theorized nearly five decades ago, and significant work has gone into understanding the physics of their formation, evolution, and stability. Several searches for T\.ZO candidates have also been carried out. Whether or not T\.ZOs could even exist or if they would be stable after formation has also been investigated. Understanding the existence and possible prevalence of T\.ZOs would have important effects on our understanding of binary evolution, stellar mergers, and inform binary population synthesis models. In this chapter, we review the formation channels, evolution and structure, final fates, and observable signatures of T\.ZOs, as well as candidates in the literature, from the inception of T\.ZO theory to recent progress in the field.
\end{abstract}

\begin{BoxTypeA}[chap1:box1]{Key Points}
\begin{itemize}
\item Thorne-\.Zytkow Objects (T\.ZOs) are an interesting avenue of binary evolution, wherein a neutron star merges with the core of a companion red supergiant. They were first theorized nearly fifty years ago, and many authors have worked to understand their evolution and whether or not T\.ZOs can exist.
\item The exact formation mechanisms of T\.ZOs are still unclear. Two promising formation avenues are a merger following common envelope evolution or a direct collision prompted by the natal kick received by the neutron star during the supernova. However, recent works have questioned how much of the T\.ZO envelope would remain bound after the merger.
\item The internal structure, nucleosynthesis, and evolution of T\.ZOs is very complex. Recent works modeling T\.ZOs predict several observable properties that are at odds with previous models, particularly for what kinds of abundance enhancements might appear on the surface of T\.ZOs. The final fates of T\.ZOs are similarly complex, with several suggested outcomes.
\item Several candidate T\.ZOs have been identified, though none have been without an alternative explanation. The strongest candidate is HV 2112, located in the Small Magellanic Cloud.
\end{itemize}
\end{BoxTypeA}

\section{Introduction}\label{ch1:intro}

Massive stars, despite their relatively short lives, play a key role in nearly every area of astrophysics, having a disproportionate impact on their environments from local to cosmological scales. It is also clear that many massive stars are born in binary systems with sufficiently short periods that binary interactions will occur throughout their life \citep[and references therein]{sana_binary_2012,marchant:23}, possibly modifying their observational properties and impact on their surroundings. To understand these binary interactions, it is important to consider all potential outcomes, including the most exotic ones.

One interesting product of binary evolution, specifically a star and compact object merger product, are Thorne-\.Zytkow Objects (T\.ZOs). First theorized by \citet{thorne:75,thorne:77} in the 1970s, they luminous red stars with neutron star (NS) cores, created through the merger of the NS with a red supergiant (RSG). While the first mentions of stars with cores made of degenerate neutrons \citep{gamow:37,landau:38,oppenheimer:38} began soon after the discovery of the particle by \citet{chadwick:32}, \citet{thorne:75,thorne:77} marked the first example of internal structure calculations. Several attempts have been made to identify T\.ZOs within our galaxy and in nearby galaxies. Interest in understanding the theory of T\.ZOs and their existence is high, as confirming this pathway of binary evolution places constraints on all other stellar merger pathways.

In this chapter, we review the history of work on T\.ZOs from their theoretical inception to today. We first review the proposed formation scenarios for T\.ZOs in \S\ref{ch2:formation}, followed by their internal structure and evolution in \S\ref{ch3:structure}. In \S\ref{ch5:finalfates} we review the final fates of T\.ZOs. Observable signals of T\.ZOs are presented in \S\ref{ch6:observables}, and previous and current T\.ZO candidates are reviewed in \S\ref{ch7:candidates}. Finally, we summarize and present a future outlook for T\.ZO research in \S\ref{ch8:summary}.

\section{Formation of Thorne-\.Zytkow Objects}\label{ch2:formation}

The original work by \cite{thorne:75, thorne:77} focused on the structure and evolution of T\.ZOs (\S\ref{ch3:structure}).
However, the question of their origin is only briefly touched upon in the final section of \cite{thorne:75}, where they speculate on their formation.
They propose that a T\.ZO could result from the aftermath of stellar collapse (\S\ref{subsec:stellar_collapse}), supercritical accretion onto a NS or the inspiral of a NS into a stellar envelope (\S\ref{subsec:binary_coalescence}).
There is overlap with the formation avenues later summarized in \cite{podsiadlowski:95}, where T\.ZO are predicted to form via collisions in dense dynamical environments or following a supernova explosion (\S\ref{subsec:collisions_and_dynamics}) or from binary evolution following a massive X-ray binary phase (\S\ref{subsec:binary_coalescence}).
The recent evidence suggests that NS progenitors are frequently found in multiple star systems \citep{sana_binary_2012,duchene_stellar_2013,2014ApJS..215...15S,moe_mind_2017,offner_2023}, enhancing the prospects of formation via binary evolution (\S\ref{subsec:binary_coalescence}), collisions and dynamical interactions (\S\ref{subsec:collisions_and_dynamics}), in contrast to T\.ZO formation during stellar collapse (\S\ref{subsec:stellar_collapse}).

The formation of T\.ZOs remains a highly active area of research, with ongoing debate about whether nature can produce the type of configurations as originally envisioned.
The assembly of T\.ZOs has been studied with a diverse range of methodologies, such as (semi)analytic calculations \citep[e.g.][]{1975MNRAS.172P..15F,1987A&A...184..164R}, stellar structure and evolution models \citep[e.g.][]{thorne:75, thorne:77,biehle:91,farmer:23,2024ApJ...971..132E}, hydrodynamic simulations \citep[e.g.][]{1978ApJ...222..269T,1995ApJ...445..367T,1996ApJ...460..801F,2023arXiv231106741H} and population synthesis \citep[e.g.][]{2018JApA...39...21H,NathanielTZOs}.
These methods are complementary and explore the complex physics that takes place in the formation of T\.ZOs, including a wide range of spatial and temporal scales, the evolutionary processes in massive stars, NS formation and structure, magnetic fields and jet formation as well as overall population statistics.
We now proceed to explore the proposed formation scenarios: stellar collapse (\S\ref{subsec:stellar_collapse}), binary coalescence (\S\ref{subsec:binary_coalescence}) as well ass collisions and dynamics (\S\ref{subsec:collisions_and_dynamics}).

\subsection{Stellar Collapse}\label{subsec:stellar_collapse}
The theory of stellar evolution predicts that stars several times more massive than the Sun will predominantly end their nuclear burning cycles with a dramatic explosion \citep[e.g.][]{2002RvMP...74.1015W,2003ApJ...591..288H}. These explosions are likely to remove most of the stellar envelope and often lead to the formation of a NS. If all of the envelope is ejected, the supernova results in the formation of a naked NS. \cite{thorne:75} speculate if the collapse of the degenerate electron core in a (super)giant could avoid the ejection of the envelope and lead to the formation of a T\.ZO.

There are three ways in which a T\.ZO could form during stellar collapse.
The first one is as a natural outcome of the supernova. 
During collapse of the stellar core to a proto-NS, there is a hydrodynamic response of the star to the sudden mass loss via neutrinos \citep{1980Ap&SS..69..115N,2013ApJ...769..109L}; this mass-decrement effect itself is likely to remove, at least partially, the hydrogen-rich envelope.
Furthermore, there is no evidence of a stellar envelope around the central region of SN 1987 A \citep{2024Sci...383..898F} nor in supernova remnant Cassiopeia A \citep{2013ApJ...777...22E}. The atmosphere of a NS could contain small amounts of hydrogen \citep{1997ApJ...491..270L}, but this amount is significantly less than a solar mass; moreover, it has a different structure than a stellar envelope.
The second one is from material that falls back onto the NS shortly after initial explosion. This fallback could carry some of the matter close to the vicinities of the NS; nonetheless, this matter would consist predominantly of heavy elements instead of hydrogen.
The third one is from accretion of gas in their surroundings, long after NS formation. However, most surrounding gas was likely ejected during the explosion that resulted in the formation of the NS and any remaining gas has densities orders of magnitude smaller than in stellar envelopes \citep[e.g.][and references therein]{2019ApJ...884...22A}.
None of these scenarios leads naturally to the canonical structure of a T\.ZO (\S\ref{ch3:structure}).
Therefore the uncertain prospects of T\.ZO formation from single stellar evolution are not favorable \citep{1996ApJ...460..801F}, but stellar companions can make a difference.

\subsection{Binary Coalescence}\label{subsec:binary_coalescence}
Originally, \cite{thorne:75} suggested that the inspiral of a NS into a stellar envelope could result in the formation of a T\.ZO.
NS progenitors are predominantly presumed to be interacting stellar binaries \citep{sana_binary_2012}.
Therefore, mass transfer episodes involving a stellar donor and a NS companion are not uncommon \citep[e.g.][]{NathanielTZOs}.
If the mass transfer episode is dynamically unstable it can lead to a \textit{common envelope} phase \citep[see][and references therein]{2013A&ARv..21...59I,2020cee..book.....I,2023LRCA....9....2R}, which is extremely complex.
During this phase, the NS is embedded in the hydrogen-rich stellar envelope and gas drag facilitates the dynamical inspiral of the inner binary, composed of the NS and the stellar core within the common envelope.
However, one plausible outcome of the common envelope phase is that the NS and the stellar companion merge.
This particular outcome has been used to estimate the number of T\.ZOs in the Galaxy today, ranging from a few \citep[e.g.][]{NathanielTZOs,2024arXiv241017315R} and up to 20-200 \citep[e.g.][]{podsiadlowski:95,2018JApA...39...21H,2024arXiv241017315R}.
However, there are many uncertainties around this binary coalescence scenario and the assembly of T\.ZOs.

The first uncertainty in the coalescence scenario is if the common envelope phase results in either a short period binary or a merger between the star and the NS \cite[e.g.][]{1978ApJ...222..269T,1995ApJ...445..367T}. The fate of the common envelope depends on several factors, including the evolutionary phase of the star at the onset of the coalescence, the response of the stellar donor to stripping throughout the phase and the hydrodynamic evolution of the inner binary within the common envelope.
Assuming the coalescence leads to a merger, another major uncertainty arises: whether or not the core of the star is disrupted.
The disruption of the core is a violent process which can release energy directly from shocks and heating.
Additionally, the formation of an accretion disk can lead to additional feedback \citep[e.g.][]{podsiadlowski:95} and potentially jets \citep{2015MNRAS.449..288P}, which could prevent the formation of the T\.ZO and could lead to the prompt formation of a (mildly) recycled pulsar or a black hole \citep{1997ApJ...478..713G}.
For example, \cite{2023arXiv231106741H,2024ApJ...971..132E} have explored the conditions of disk formation in T\.ZO, finding that accretion disks generally form and potentially prevent \textit{conventional} T\.ZOs to be formed.
\cite{2024ApJ...971..132E} suggest that a slightly different configuration with a \textit{thin-envelope} (\S\ref{ch5:finalfates}) may be a more natural outcome from binary coalescence.

There are some speculative alternatives that slightly deviate from the standard coalescence scenario, which we proceed to mention for completion.
\cite{1975MNRAS.172P..15F} proposed that close star+NS binaries can be formed in globular clusters via tidal capture, a configuration that could then follow the coalescence scenario as described above (but see \S\ref{subsec:collisions_and_dynamics}).
\cite{2022MNRAS.513.4802A} suggest an alternative formation scenario for T\.ZOs: the stellar core-merger-induced collapse. This scenario involves the merger of a ONeMg white dwarf with the stellar core of a massive star, which would then collapse the white dwarf into a NS. \cite{2022MNRAS.513.4802A} suggests that the energy released by nuclear fusion in an O+Ne deflagration will not unbind the core. This formation avenue, in principle, could avoid the core disruption prompted during the merger of a more compact object, such as a NS or a black hole, and could lead to the formation of a millisecond pulsar \citep{1990ApJ...353..159B}.
Finally, a T\.ZO can form in triple and quadruple star systems (see also \S\ref{subsec:collisions_and_dynamics}).
\cite{eisner_planet_2022} suggest that the recent observed massive compact triple TIC 470710327 could eventually merge and reduce to a massive stellar binary, similar to the ones we described will coalesce. The stellar multiplicity reduction in this formation scenario would be a direct outcome of stellar evolution rather than dynamics (\S\ref{subsec:collisions_and_dynamics}).

\subsection{Collisions and Dynamics}\label{subsec:collisions_and_dynamics}
The presence of NSs in Galactic globular clusters suggests that dense dynamical environments can host collisions between NSs and non-degenerate stars \cite[see][and references therein]{2002ApJ...573..283P}.
Collisions have been proposed as a formation avenue for T\.ZOs, either as a plausible outcome of tidal captures \citep{1975MNRAS.172P..15F,1987A&A...184..164R} or collisions \citep{1992ApJ...389..546B} in dense dynamical environments, or from collisions following supernovae in binaries \citep{1994ApJ...423L..19L,1995MNRAS.274..461B,hirai_neutron_2022}.
Generally, collisions and dynamics can lead to the formation of a close binary embedded in a common envelope \citep{1975MNRAS.172P..15F,1987A&A...184..164R}, more in line to the coalescence scenario.
Alternatively, direct collisions between a NS and its stellar companion could facilitate the formation of T\.ZOs; the range of conditions under which these direct collisions occur is quite limited \citep{1987A&A...184..164R,hirai_neutron_2022}.
While the formation rates of T\.ZOs are uncertain, the binary coalescence scenario has often been considered to dominate the formation rate of T\.ZOs \citep{1994ApJ...423L..19L,podsiadlowski:95}.

Alternatively, dynamics can also prompt collisions and mergers in multiple-star systems. In multiple-star systems, dynamical instabilities can arise at different evolutionary stages. These instabilities can ultimately lead to a tidal capture or collision similar to those mentioned above, but emerging from a different initial configuration and environment. There are some peculiar configurations that could result in close encounters between a NS and a non-degenerate star. One of them may occur in certain triple star configurations: the von-Zeipel-Kozai-Lidov mechanism \citep{1910AN....183..345V,1962AJ.....67..591K,1962P&SS....9..719L}. This mechanism can arise when an inclined binary is perturbed by a third body, resulting in an oscillation between the eccentricity of the inner binary and the inclination of the binary with respect to the total angular momentum vector. The variation in eccentricity results in a decrease in the separation which can result in the merger of the inner binary. If the inner binary is comprised of a NS and a non-degenerate star, this could be the progenitor of a T\.ZO. Additionally, a wide ($\sim 10^3-10^4$ au) NS and non-degenerate star binary can be similarly affected by the Galactic tide \citep{Stegmann_2024}. However, these wide binary configurations are sensitive to the NS natal kick, which can drastically affect the structure of the system. Moreover, the same caveats from the binary coalescence and collision scenario hold.

\section{Structure and evolution of Thorne-\.Zytkow Objects}
\label{ch3:structure}

\subsection{Brief history of numerical models}

As mentioned in \S\ref{ch1:intro}, while the idea of stars with a ``neutron core''
dates from the 1930s, the first calculation for the internal
\emph{structure} were presented in \cite{thorne:75, thorne:77}. These
models assumed thermal and hydrostatic equilibrium (bypassing the
formation question, see \S\ref{ch2:formation}) and were constructed
combining \cite{paczynski:69, paczynski:70}'s stellar structure code
for the outer envelope with ad-hoc numerical tools for the innermost
layers, accounting for general relativistic effects. The
\emph{evolution} was discussed in terms of connecting sequences of
structures, but not directly computed.
 
Since then, few authors if any have re-considered the innermost
structure next to the NS core, and most subsequent studies have
focused on predicting observable properties for the outer envelope
(with some notable exceptions, e.g.,~\citealt{zimmermann:79phd,
  bisnovaty:84, eich:89}). Interests in these exotic theoretical
stellar objects renewed in the early 1990s \citep{biehle:91,cannon:92, cannon:93,biehle:94, podsiadlowski:95}. This enabled the use of stellar \emph{evolution} codes to self-consistently step through equilibrium structures, modeling the neutron star and its immediate surroundings with a custom inner boundary condition for the stellar model. In the mid 2010s, observational claims (see also \S\ref{ch6:observables})
lead to more theoretical and modeling efforts \citep[most recently by][]{farmer:23}.

Since the pioneering work of \cite{thorne:75, thorne:77}, the
computational strategy has remained similar \citep{farmer:23}, and is
based on separating the calculation of the structure ($L$, $\rho$,
$T$, etc. as a function of $m$ or $r$) and the composition and energy
generation (``operator split''). The structure can be calculated
assuming the energy generation to be confined to a very small inner
region, producing a family of structures for a given inner boundary.
These can then be matched to composition structures generating
sufficient energy to sustain the structure to construct, when possible,
self-consistent models.

The most important updates in the past $\sim$50\,years involve the
treatment of nuclear burning (from a CNO nuclear network designed for
main sequence burning in the 1970s, to progressively larger nuclear
reaction networks accounting for the finite time for $\beta$-decays,
and ultimately the ``interrupted rapid proton capture'' process,
\citealt{wallace:81}) and the treatment of opacities and neutrino
cooling. But maybe most importantly, progress in computing power has
allowed for dramatic increases in spatial and temporal resolution, and
for on-the-fly treatment of the time evolution of these structures:
for example, the entire lifetime of $\sim{}10^{5}$\,years for
\cite{farmer:23} models corresponds to only one single timestep of
\cite{cannon:92, cannon:93}.

All combined, these updates have caused changes in the details of
theoretical prediction and re-opened many questions already raised in
\cite{thorne:75, thorne:77}. These include (\emph{i}) what is the
source of energy sustaining these structures? (\emph{ii}) what
processes determine the opacity structure, or in other words: do the
inner layers become hot enough to produce electron-positron pairs? and
(\emph{iii}) are these structures dynamically stable? Below, we
attempt to summarize the current status for each of these questions.

\subsection{Overview of the internal structure}

\begin{figure}[htbp]
  \centering
  \includegraphics[width=0.25\textwidth]{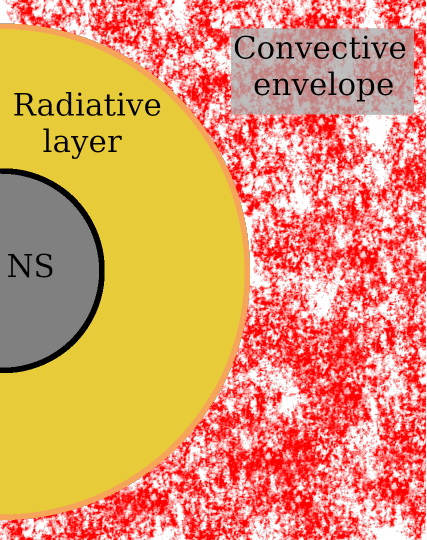}
  \caption{Inner structure of \tzo~ (not to scale). 
    Different authors include in
    their computational domains different fraction of the layers
    depicted. See also Fig.~1 in \cite{thorne:77}.
    }
  \label{fig:cartoon}
\end{figure}

\cite{thorne:75,thorne:77} built their models by analogy with red
giant stars with cores sustained by degenerate electron gas, that is
approximately with a white dwarf core instead of a NS core (see
Fig.~\ref{fig:cartoon}). They envisioned the inner
$\sim10\,\mathrm{km}$ making the NS core surrounded by a thin layer
where matter is non-degenerate and energy is efficiently carried by
radiation (so called ``halo''). However, moving outwards in radius $r$
(or equivalently mass coordinate $m$), at some point radiation will
not be able to carry all the energy flux and convection must set in.
This is because in the inner regions the opacity $\kappa$ is dominated
by electron scattering, but as the temperature decreases, the
relativistic correction to the electron opacity become less and less
important, increasing $\kappa$ and thus decreasing the ``critical
luminosity'' \citep[e.g.,][]{joss:73, paxton:13}
\begin{equation}
  \label{eq:L_crit}
  L_\mathrm{crit} \equiv L_\mathrm{crit}(r) \simeq \frac{4 \pi G c m}{\kappa} \ \ ,
\end{equation}
where $m\equiv m(r)$ is the Lagrangian mass coordinate,
$\kappa\equiv\kappa(r)$ is the \emph{local} opacity, $c$ is the speed
of light, and $G$ the gravitational constant. $L_\mathrm{crit}$ in
Eq.~\ref{eq:L_crit} is an ``effective'' Eddington luminosity
accounting for the \emph{local} opacity $\kappa\equiv\kappa(r)$ (as
opposed to the electron scattering opacity), and it neglects general
relativistic corrections, which can modify $L_\mathrm{crit}$ by
$\lesssim 30\%$, \citep{thorne:75}. Wherever
$L(r) \geq L_\mathrm{crit}(r)$ radiation is too impaired to
efficiently carry energy, and convection necessarily sets in
\citep[e.g.][]{joss:73, paxton:13}. From this point, commonly
referred to as the ``knee'' (orange line surrounding the yellow radiative layer in Fig.~\ref{fig:cartoon}),
onwards the envelope is fully convective, and the temperature-density
relation steepens significantly (see, e.g., Fig.~1 in
\citealt{thorne:75}).

\begin{figure}[htbp]
  \centering
  \includegraphics[width=0.5\textwidth]{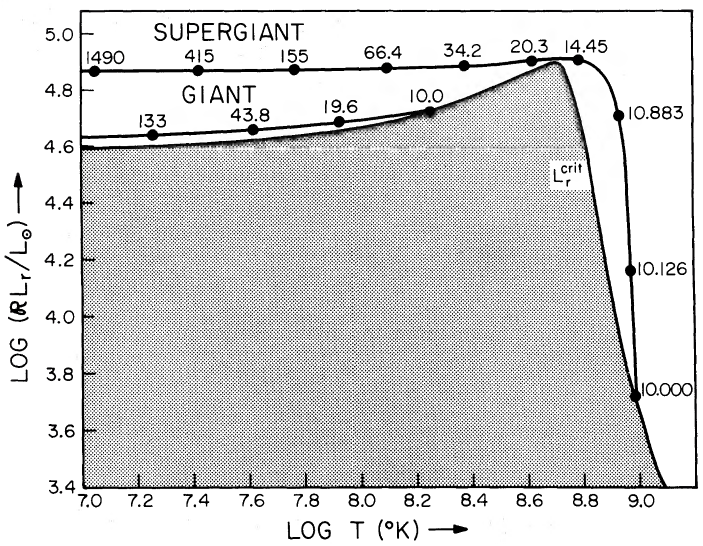}
  \vspace*{5pt}\includegraphics[width=0.495\textwidth]{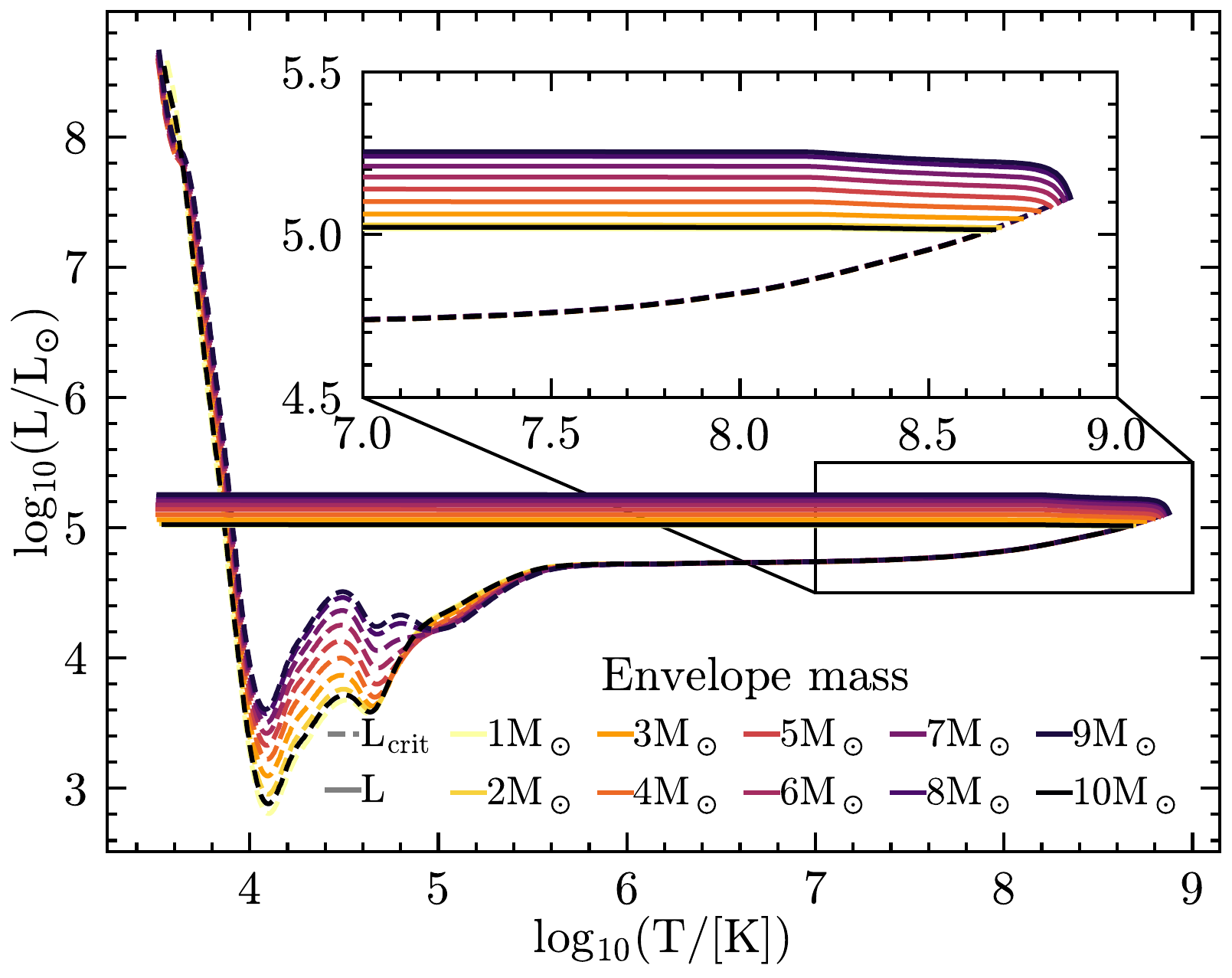}
  \caption{Luminosity and critical luminosity profiles as a function
    of temperature. The left panel shows models of total mass
    5\,$M_\odot$ (giant) and 12\,$M_\odot$ (supergiant) from
    \cite{thorne:77} (see their Fig.~4), with permission. The right panel shows models
    from \cite{farmer:23}, which assume a $1.4\,M_\odot$ NS in the
    core and are labeled by the total initial envelope mass. The left
    panel only shows a range of temperatures comparable to the inset
    panel on the right, includes relativistic corrections (redshift
    factor $\mathcal{R}$, neglected in the right panel), and labels
    the radius of each mesh points in km (numbers along the solid
    lines representing $\log_{10}(\mathcal{R}L/L_\odot)$).
    \cite{farmer:23} computed only the structure above the ``knee''
    where $L(r) = L_\mathrm{crit}(r)$ and are plotted after $10^4$
    timesteps of evolution, corresponding to a representative point
    during the evolution. \cite{farmer:23} models are publicly
    available at on Zenodo at \citet{farmer_2023_4534425}.}
  \label{fig:L_Lcrit}
\end{figure}

Fig.~\ref{fig:L_Lcrit} shows the luminosity and critical luminosity
profiles for the original models from \cite{thorne:75, thorne:77}
(left panel) and models from \cite{farmer:23} (right panel), using the
temperature as independent coordinate. It shows that from the
innermost boundary until the surface (where $\kappa$ sharply
decreases, causing an increase in $L_\mathrm{crit}$, shown only in the
right panel) these model need to be convective. Moreover, the figure
also shows how $L$ is approximately constant in these models: most of
the energy is generated in the inner most layer where $L$ raises
sharply, this validates a posteriori the operator-split approach.

\subsubsection{Gravitational- vs.~nuclear-powered \tzo s}

In this section, we address the energy generation question (\emph{i})
introduced above. \cite{thorne:75, thorne:77} found two families of
self-consistent solutions, which they named ``giant'' and
``super-giant'' \tzo s, respectively. The left panel of
Fig.~\ref{fig:L_Lcrit} shows two representative examples. The
``giant'' solutions are fully powered by the release of gravitational
energy from the accretion of the envelope onto the ``halo''
surrounding the NS. For these models, $L<L_\mathrm{crit}$ at a certain
radius below which all the energy is generated: any nuclear processing
happens in a radiative region below the ``knee'' where convection sets
in, and the products of nuclear burning are not expected to be mixed
outwards to the surface. Therefore, no chemical signature is expected,
and distinguishing them from red giant stars with electron-degenerate
cores may be extremely challenging (\S\ref{ch6:observables}).

Increasing the mass of the envelope (that is, for fixed NS mass,
increasing the total mass), \cite{thorne:75, thorne:77} first find a
gap where no self-consistent solution for the structure and energy
generation can be found, and above that, the ``super-giant'' solutions
appear. In these models, the vast majority of the energy is provided
by nuclear burning which happens in the convective region itself since
$L\geq L_\mathrm{crit}$ throughout the computational domain. This
suggests that potential peculiar chemical signatures of burning may be
advected to the surface by convection and lead to observable
signatures (\S\ref{sec:nuclear_challenges}).

This illustrates the two possible power sources for \tzo s, but also
introduces one of the open questions remaining: is the separation of
solutions into two families (gravitationally powered ``giants'' and
nuclear powered ``supergiants'') physical or a numerical artifact? In
fact, different authors using different codes and approximations have
either confirmed \citep[e.g.][]{biehle:91, cannon:93} or questioned
\citep[e.g.][]{cannon:92, farmer:23} this result. For instance, the
most recent models of \cite{farmer:23}, which only compute the
structure \emph{above} the knee, do not find a mass (or equivalently,
luminosity) gap and two families of solutions -- suggesting that
numerical details and uncertain stellar physics processes (e.g.
convective boundary mixing) may play an important and as-yet not
understood role. The numerical complexity of these models makes
exploring the systematic uncertainties due to computational approach
and physical unknowns a challenging \citep{farmer:23}, but still
necessary, task.

\subsubsection{Supergiant \tzo~nuclear challenges and signatures}
\label{sec:nuclear_challenges}

Because of the opportunity for nuclear/chemical signatures in
``super-giant'' nuclear-powered \tzo s, these have received much more
theoretical attention. These models are supported by the energy
generation via nuclear fusion which happens at the bottom of a
convective layer connected to the photosphere by convective mixing.
This, on the one hand, creates the opportunity for observing nuclear
signatures at the surface, but on the other hand challenges the
typical approach to nuclear energy generation in models of stars.

\begin{figure}[htbp]
  \centering
  \includegraphics[width=0.5\textwidth]{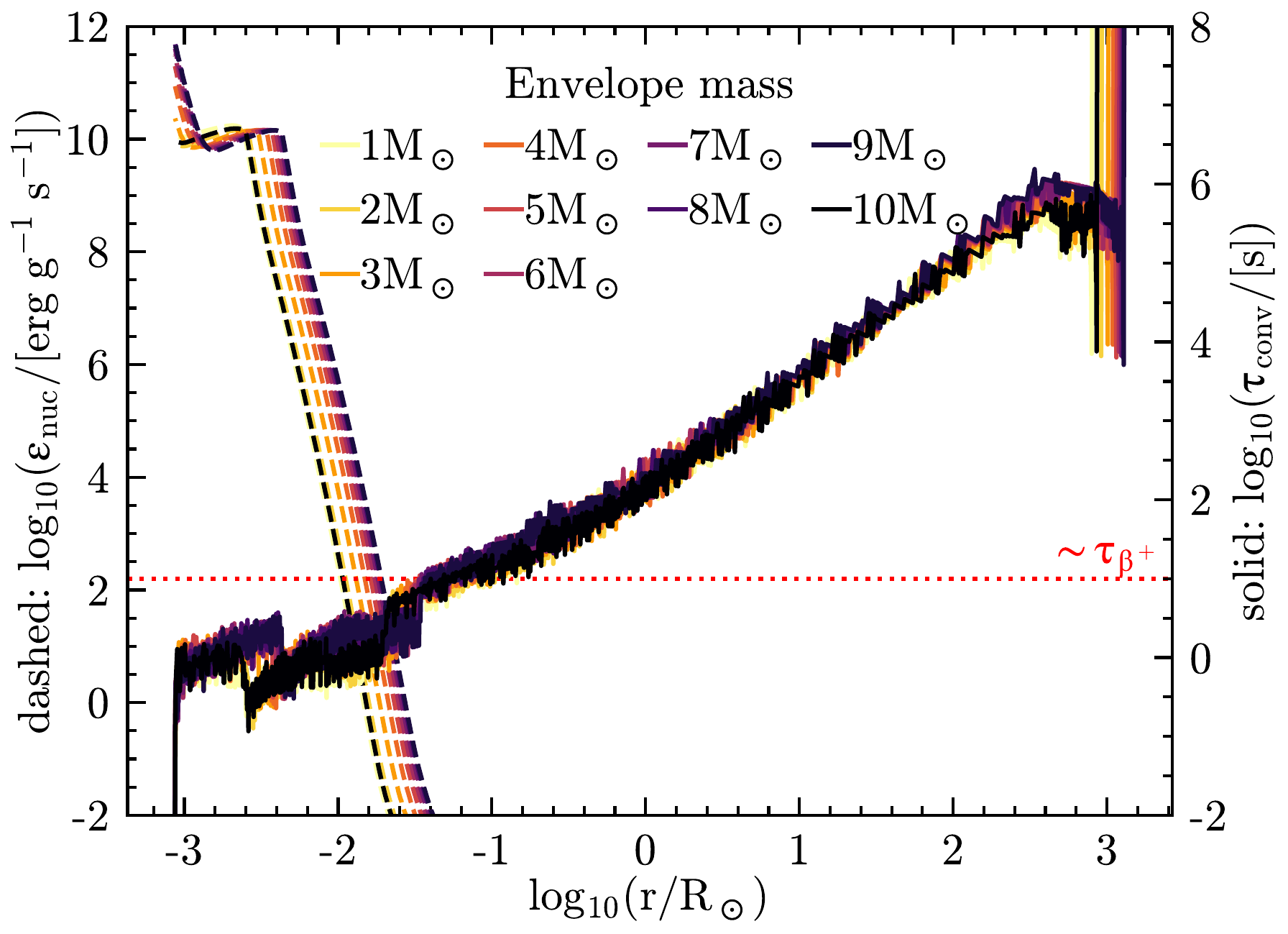}
  \caption{Local convective turnover timescale (solid lines, right
    y-axis) and energy generation rate from nuclear reactions (dashed,
    left y-axis) for \cite{farmer:23, farmer_2023_4534425} models.
    The dotted red horizontal line marks the approximate half-life
    $\tau_{\beta^{+}}$ for $^{13}\mathrm{N}$ and $^{15}\mathrm{O}$,
    which is longer than the convective turnover timescale in the
    region where most energy is released by the nuclear reactions. The
    convective turnover timescale profile is noisy because it is
    obtained in post-processing and depends on the mesh details.}
  \label{fig:tau_conv}
\end{figure}

The original models from \cite{thorne:75, thorne:77} admittedly used
an unsatisfactory nuclear reaction network only capable to simulate
the main-sequence CNO cycle \citep{bethe:39}. This was already known
to be insufficient since the CNO-cycle involves $\beta^{+}$ decays
$^{13}\mathrm{N}(e^+, \nu_e)^{13}\mathrm{C}$ and
$^{15}\mathrm{O}(e^+, \nu_e)^{15}\mathrm{N}$, where the parent nuclei
($^{13}\mathrm{N}$ and $^{15}\mathrm{O}$) have half-lives
$\tau_{\beta^{+}}\lesssim 10\,\mathrm{sec}$, which is comparable to or
longer than the convective turnover timescale at the base of the
convective envelope where the nuclear burning happens
\citep{thorne:75, thorne:77, zimmermann:79phd, biehle:91, cannon:92,
  cannon:93, farmer:23}. The convective turnover timescale relevant to
the composition in a given burning region can be estimated as
\begin{equation}
  \label{eq:conv_turnover}
  \tau_\mathrm{conv} \equiv \tau_\mathrm{conv}(r) = \Delta r/v_\mathrm{conv} \ \ ,
\end{equation}
with $\Delta r$ the local mesh size in radius and $v_\mathrm{conv}$
the local convective velocity. Fig.~\ref{fig:tau_conv} shows the
$\tau_\mathrm{conv}$ (right y-axis) and nuclear energy generation
profiles (left y-axis) for models from \cite{farmer:23}. Since the
local convective turnover timescale is shorter than
$\tau_{\beta^{+}}$, the unstable isotopes $^{13}\mathrm{N}$ and
$^{15}\mathrm{O}$ can be mixed out of the region sufficiently hot for
burning before they decay, making their daughter nuclei unavailable to
complete the CNO cycle. One of the consequences is that the energy
release by the $\beta^{+}$-decays is also spread throughout the
convective envelope (hence the relatively slower decay of the dashed
lines in Fig.~\ref{fig:tau_conv} compared to main-sequence massive
stars).

This problem was already identified in \cite{thorne:75, thorne:77},
and subsequent work from \cite{zimmermann:79phd, eich:89, biehle:91,
  cannon:93, farmer:23} focused on improving the numerical treatment
of nuclear burning. Specifically, \cite{eich:89, biehle:91} first
identified that the physical conditions in a \tzo~will lead to the
interrupted rapid proton capture process \citep{wallace:81} generating
most of the energy. In fact, the envelope is proton-rich, and proton
captures on seed nuclei at the hot temperatures at the envelope base
dominate the energy release. The proton capture process however is
continuously interrupted by convective mixing advecting the seed
nuclei outwards in cooler (and thus non-burning) regions of the
envelope, where unstable proton-rich nuclei can decay depositing part
of the energy, and changing the type of seed nuclei encountered by the
large proton flux at the base of the envelope at the next convective
turnover.

This ``interruption'' of the proton captures by convection leads to a
change in the potentially observable composition at the photosphere,
and a change in the hydrogen mass fraction and thus opacity $\kappa$.
The first effect may lead to observable signatures: \cite{biehle:91}
proposed $^{84}\mathrm{SrH}$ as a potentially observable molecule that
may form at the surface from the binding of $^{84}\mathrm{Sr}$, one of
the nuclei produced by subsequent proton captures in the burning
region, with hydrogen. Other classical signatures that have been
proposed (and looked for in observed samples of RSGs) are the presence
of Calcium \citep[e.g.,][]{biehle:91, farmer:23}, Rubidium, Strontium,
Yttrium, and Molibdenum \citep[e.g.,][]{cannon:93}. Another notable
element that may be overproduced in \tzo s is Lithium through the
\cite{cameron:71} process
$^3\mathrm{He}(\alpha,\gamma)^7\mathrm{Be}(e^{-},
\bar{\nu}_e)^7\mathrm{Li}$, as proposed by \cite{podsiadlowski:95}.
While $^7\mathrm{Li}$ is notoriously easy to destroy in the burning
regions of any star, advection of $^7\mathrm{Be}$ out of the burning
region may lead to its $\beta^{-}$ decay into $^7\mathrm{Li}$ outside
of the region hot enough for burning and make the surface Li-rich.

However, all these nucleosynthetic signatures are unfortunately
\emph{not} unique to \tzo s \citep{tout:14, maccarone:16, farmer:23}.
The main known ``false positive \tzo'' based on these signatures are
super-asymptotic giant branch (AGB) stars \citep{tout:14, doherty:17}
where nucleosynthesis is powered by the s-process \citep{kuchner:02},
or stars polluted by an AGB companion before they themselves became
AGB stars \citep{maccarone:16}. Given these should vastly outnumber
\tzo s in any stellar population \citep{NathanielTZOs}, it does not
seem promising to identify candidates based on \emph{stable heavy
  isotopes} in observed AGBs and RSGs.

A possible alternative strategy to identify \tzo~through
nucleosynthetic signatures proposed by \cite{farmer:23} is to rely on
\emph{unstable} isotopes with lifetimes short compared to stellar
lifetimes (of the \tzo~candidate and/or possible ``false positive
\tzo''). These can also be produced deep in the envelope and advected
to the surface by convection, and because of their short lifetime,
they have to be continously replenished inside the star to be
observable. One such unstable isotope is $^{44}\mathrm{Ti}$ which has
a $\sim60\,\mathrm{yr}$ half-life \citep{ahmad:06} and may be observed
bound to oxygen in $\mathrm{TiO}$ molecules: observing
$^{44}\mathrm{TiO}$ or $^{44}\mathrm{TiO}_2$ molecules would strongly
imply an in-situ production of $^{44}\mathrm{Ti}$ which can only
happen beyond silicon core burning or during stellar explosions
\citep{timmes:96}, but can be sustained for the entire lifetime of a
``supergiant'' \tzo. Note that contamination from the nucleosynthesis
in the NS progenitor should be easy to rule out, since the half-life
of $^{44}\mathrm{Ti}$ is very short compared to thelifetime of
supernova remnants \citep{farmer:23}.

\subsection{Stability and evolution}

The evolution of supergiant \tzo s can be driven by the change in
composition (and thus opacity) in the envelope due to the nuclear
burning, the (notoriously uncertain, \citealt{smith:14, renzo:17,
  beasor:20, decin:24}) mass-loss from the envelope, or the onset of
dynamical instabilities \citep[e.g.,][]{farmer:23}. Because of the
numerical complexity of \tzo~models, few authors have computed their
evolution until the end of their lifetime, but interestingly all
estimates of lifetimes (regardless of what limits them) are between
$\sim 10^{5}-10^{6}\,\mathrm{years}$.

In the most recent models from \cite{farmer:23}, the onset of
dynamical instabilities in the envelope, with pulsations with growing
amplitude driving high mass-loss rates \citep[e.g.][]{yoon:10} and
eventually driving shocks that may remove the remaining envelope.
These pulsations are triggered by the changes in hydrogen mass
fraction in the envelope, which affect directly $\kappa$ throughout
the envelope, and thus $L_\mathrm{crit}$ at the ``knee''
(e.g., see Eq.~5.2 in \citealt{thorne:77}). Interestingly,
\cite{farmer:23} finds that the instability makes \emph{lower} mass
\tzo~have \emph{shorter} lifetimes. The pulsational properties of
\cite{farmer:23}'s model are also at odds with the observed
variability for claimed \tzo-candidates (\S\ref{ch7:candidates}).

\section{The Final Fate of Thorne-\.Zytkow Objects}\label{ch5:finalfates}
The life of a star as a T\.ZO can come to an end in several ways. First, mass loss from a T\.ZO can make its mass too low to sustain the structure of the T\.ZO. If there is a mass gap between the low-mass and high-mass T\.ZOs as found in classical studies \citep{thorne:77}, the mass of a massive T\.ZO can be reduced below the minimum mass to sustain its structure through mass loss. This can lead to the collapse of the T\.ZO \citep{podsiadlowski:95}. Even if there is no mass gap to have a stable T\.ZO structure \citep[e.g.,][]{farmer:23}, mass loss can eventually take all the envelope and a naked NS may be left after the T\.ZO phase. \citet{farmer:23} found that T\.ZO envelopes are dynamically unstable to pulsations, and they can experience strong mass loss caused by the dynamical instability as also suggested for RSGs \citep{yoon:10}. Another possible end of a T\.ZO is the collapse of the central region triggered by the cease of the nuclear reactions sustaining the T\.ZO structure caused by the exhaustion of the burning elements \citep{podsiadlowski:95}, although this possibility is recently suggested to be less plausible \citep{farmer:23}.

\subsection{Naked neutron stars}
If mass loss from a T\.ZO becomes significant and if it can avoid collapsing, mass loss would eventually strip the entire envelope of the T\.ZO and a naked NS can remain \citep{farmer:23}. Because the pulsation-driven mass-loss rate can be as high as $10^{-2}~\mathrm{M_\odot~yr^{-1}}$ \citep{yoon:10}, T\.ZOs with the $1-10~\mathrm{M_\odot}$ envelope can end up to be naked NSs on a timescale as short as $100-1000$~years. The properties of the post-T\.ZO NS has not been investigated and they are still not clear. The estimated accretion onto the NS during the T\.ZO phase is small and the NS mass is not likely to increase much during the T\.ZO phase. While the accretion can provide the angular momentum to spin up the NS \citep{podsiadlowski:95}, it is also possible that the NS spin down during the T\.ZO phase due to the magnetic braking caused by the interaction between the neutron-star magnetic field and the T\.ZO envelope. In the later case, a T\.ZO may end up with a slowly rotating NS surrounded by a CSM that is lost during the T\.ZO phase that can be observed as SN remnants like RWC~103 \citep{liu:15}.

\subsection{Collapse and its possible outcomes}
Several consequences are proposed after the possible collapse of T\.ZOs \citep{podsiadlowski:95}. Once the collapse is triggered, the central region heats up and the region just above the neutron-star core becomes hot enough ($\gtrsim 2.5\times 10^{9}~\mathrm{K}$) for neutrino emission to be dominant. Then the accretion can be super-Eddington and a runaway accretion towards the central core is achieved. The left panel of Figure~\ref{fig:fate} shows the free-fall accretion rate of the $16~\mathrm{M_\odot}$ T\.ZO computed by \citet{biehle:91}. The accretion rate is sustained high enough to keep the super-Eddington accretion ($\gtrsim 10^{-3}~\mathrm{M_\odot~yr^{-1}}$, \citealt{chevalier:89}). If the accretion continues with the free-fall timescale, a black hole is formed in a few months. Depending on the angular momentum of the envelope, an accretion disk can be eventually formed and the accretion timescale could be determined by the viscous timescale.

\subsubsection{Explosions}
\begin{figure}[htbp]
  \centering
  \includegraphics[width=0.5\textwidth]{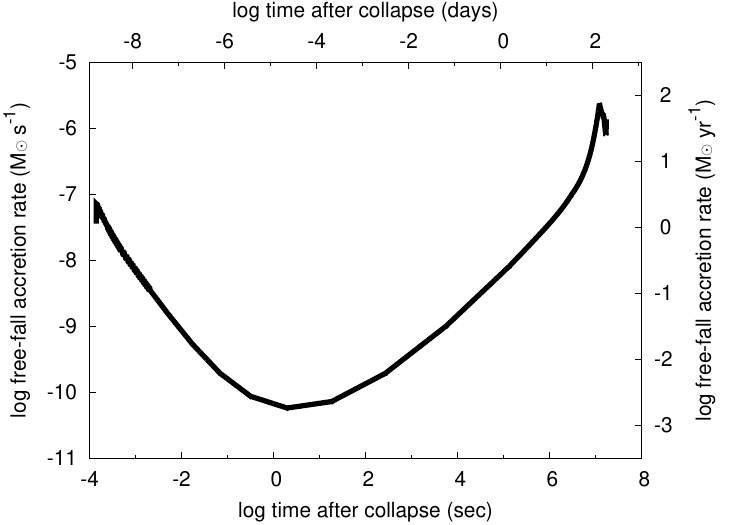}
  \vspace*{5pt}\includegraphics[width=0.495\textwidth]{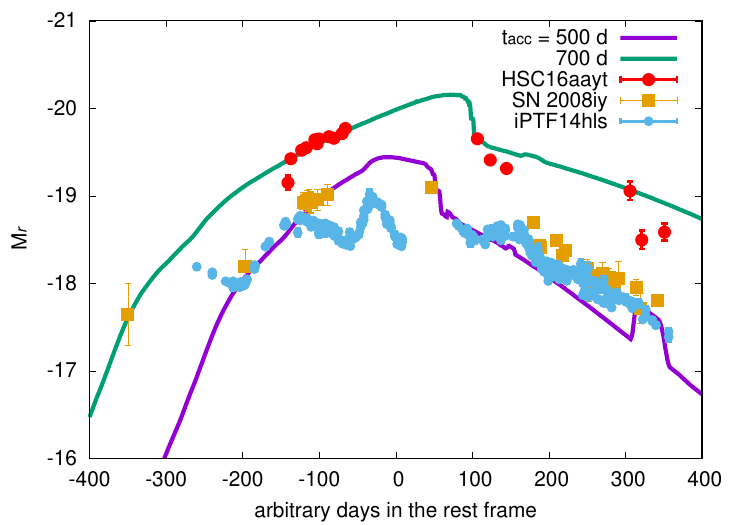}
  \caption{
  \textit{Left:} Free-fall accretion rate of a $16~\mathrm{M_\odot}$ T\.ZO from \citet{biehle:91}. This panel is adopted from \citet{moriya:18}, with permission.
  \textit{Right:} Predicted light curves of possible explosions of the $16~\mathrm{M_\odot}$ T\.ZO after the collapse. The light curve duration and luminosity are affected by the accretion timescale ($t_\mathrm{acc}$). Three long-lasting transients (HSC16aayt from \citealt{moriya:19}, SN~2008iy from \citealt{miller:10}, and iPTF14hls from \citealt{arcavi:17,sollerman:19}) are presented for comparison. This panel is adopted from \citet{moriya:21}, with permission.
  }
  \label{fig:fate}
\end{figure}

The accretion disk around the NS or black hole can trigger a large-scale outflow or jet. Such an outflow or jet can push back the accreting materials and the collapsing T\.ZO may gain enough energy to explode \citep{moriya:18}. Depending on the accretion timescale, such an explosion could be observed as a long-lasting SN (\citealt{moriya:21}, Figure~\ref{fig:fate}). The long-lasting accretion sustains for very long time, the T\.ZO collapse could be observed as an extremely long-lasting gamma-ray burst as proposed for Swift 1644+57 \citep{quataert:12}. Because T\.ZOs may have high mass-loss rates, their explosions could be affected by the dense surrounding material and they might be observed as Type~IIn SNe showing strong signatures of the interaction between the ejecta and circumstellar matter.

\cite{2023arXiv231106741H,2024ApJ...971..132E} have suggested an unstable post-collapse configuration with a plausible long configuration they refer to as thin-envelope T\.ZOs. In this configuration, a NS collapses into a black hole, produce an ultra-long gamma-ray burst and in the aftermath is able to retain less than 1\% of the initial envelope mass as an optically-thick thin-envelope. This configuration is hypothesized to be visible in X-ray, but its nature remains speculative.

\subsubsection{Pulsars and black holes with planets and low-mass stars}
If the outflow or jet is not launched, the T\.ZO envelope can end up forming a long-lasting accretion disk. The accretion can spin up the central NS and, if the central accreting NS is magnetized, strong dipole radiation that dissipates the accretion disk may be formed. In this case, a pulsar may remain after the T\.ZO collapse. If the dissipation time is long enough, planets or low-mass stars may be formed within the disk before the disk dissipation \citep{podsiadlowski:95}. Then, T\.ZOs can result in pulsars with surrounding planets or low-mass X-ray binaries.

The accretion towards the central NS may eventually end up with the formation of a black hole. In such a case, planets could be formed around the black hole like Edmunds in the movie \textit{Interstellar}\footnote{A film that Kip Thorne contributed greatly to the science of.}. If a low-mass star is formed around the black hole, it can be eventually observed as X-ray transients like V404 Cygni \citep{podsiadlowski:95}.

\section{Observable Properties of Thorne-\.Zytkow Objects}\label{ch6:observables}

In order to bring T\.ZOs out of the realm of ``fun speculation''\footnote{From \citet{thorne:75}: ``Undoubtedly the strongest reason to believe that stars with neutron cores exist in nature is the universal law that ``everything not forbidden is compulsory'' (White 1939). By comparison, any other basis for discussing existence is extremely uncertain. Nevertheless, it is fun to speculate.''}, significant work has gone into understanding how to both detect and confirm the existence of T\.ZOs. Direct detection and confirmation of a T\.ZO (or population thereof) would place extremely strong constraints not only on the stability and formation mechanisms of T\.ZOs themselves, but also on relative rates of other avenues of binary evolution. In this Section we review the observable properties of T\.ZOs and how useful those properties are for detection.

\subsection{Photometrically Derived Observables}

\subsubsection{Luminosity and Temperature}

 Although the core structure of T\.ZOs differ dramatically from `normal' RSGs, the thick convective envelopes of T\.ZOs conceal these differences from direct detection. Thus from a photometric perspective, T\.ZOs are expected to closely resemble RSGs \citep{thorne:77,farmer:23}. The original models of \citet{thorne:77} appear as very late type M supergiants, with luminosity ranges of $L\approx3\times10^{4}\text{--}1.3\times10^{5}$\ \lsol, T$_{\mathrm{eff}}\approx$2600--3100 K, $R\approx$1000\ \rsol. The updated models of \citet{cannon:92} have slightly higher luminosities at $L\approx7\times10^{4}\text{--}1.4\times10^{5}$\ \lsol.  These luminosities are consistent with those of RSGs \citep{massey:09,bonanos:09,bonanos:10,levesque:17}. While T\.ZOs are expected to be among the coldest RSGs, if not ever-so-slightly cooler \citep[$\Delta$log T$_{\rm{photosphere}}<<$ 0.1,][]{thorne:77} than a RSG at an equivalent luminosity, this \teff difference is so small that it cannot be disentangled from sources of photometric uncertainty (e.g. inter- or circumstellar dust, model uncertainties \citep{levesque:06,levesque:07ii}, instrument noise).

\begin{figure}[htbp]
  \centering
  \includegraphics[width=0.75\textwidth]{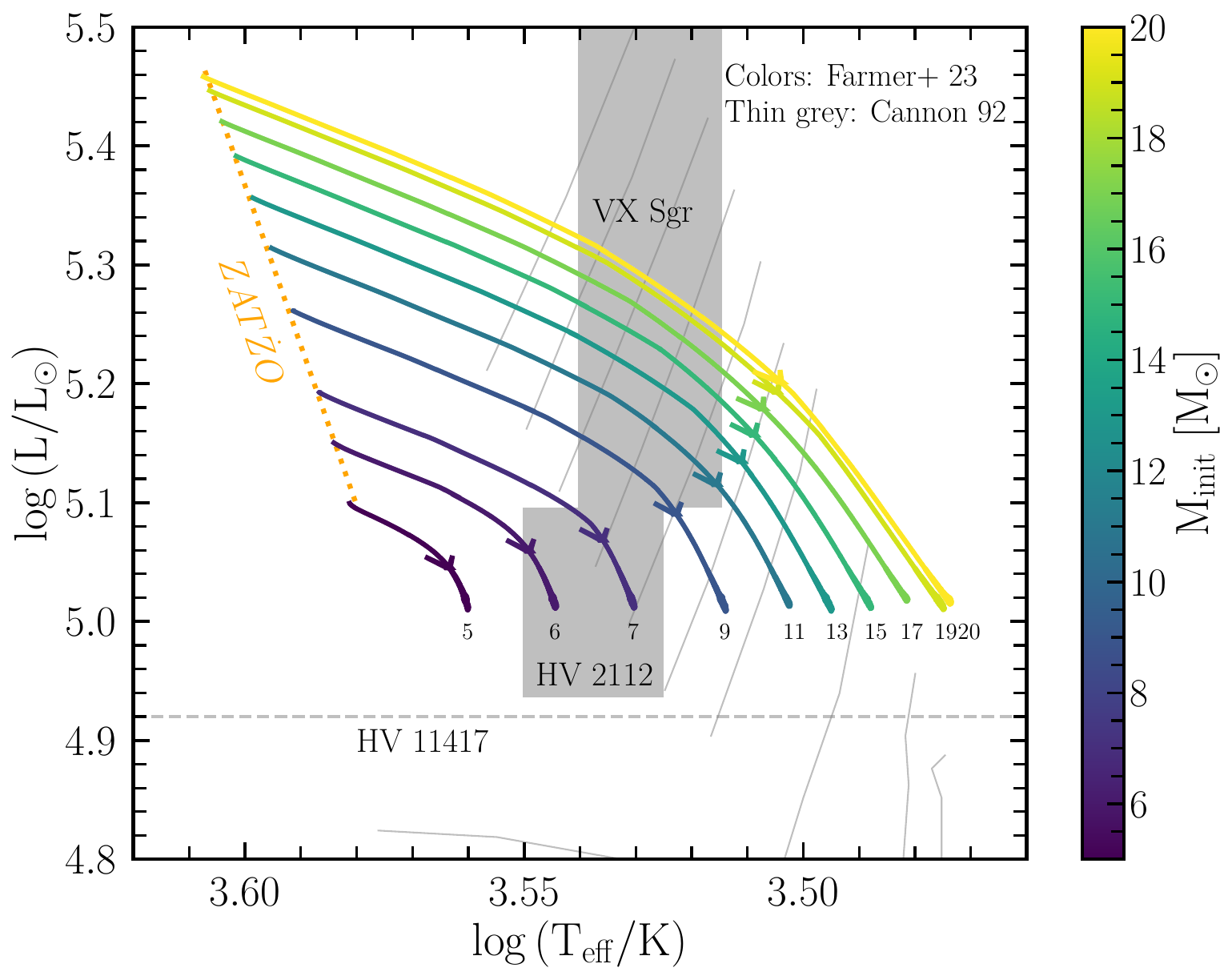}
  \caption{A Hertzsprung-Russell diagram displaying T\.ZO models from \citet[][thin grey lines]{cannon:92} and \citet[][thick colored lines]{farmer:23}. Numbers beneath each line indicate the initial mass of the \citet{farmer:23} models, and arrows indicate the direction of evolution towards decreasing luminosity. The \citet{cannon:92} tracks always evolve towards increasing luminosity. Areas corresponding to the T\.ZO candidates HV 1221, HV 11417, and VX Sgr are highlighted (see also \S\ref{ch7:candidates}).
    This figure is from \cite{farmer:23}, with permission.}
  \label{fig:farmer23hrd}
\end{figure}

The models of \citet{farmer:23}  have an overall larger range of luminosities and temperatures -- $L\approx7.1\times10^{4}\text{--}3.2\times10^{5}$\ \lsol, \teff$\approx$2800-4500 K --,  though the minimum luminosity and temperature are higher than those of  \citet{thorne:77,cannon:92}. Figure \ref{fig:farmer23hrd} displays both the \citet{cannon:92} and \citet{farmer:23} models on a Hertzsprung-Russell diagram. These luminosities and temperatures are greatly affected by not only the mass of the T\.ZO, but the initial metallicity (Z) and helium (Y) fraction as well. At higher Z, \teff and $L$ decrease; at higher Y, \teff and $L$ increase.

Therefore, luminosity and temperature alone are not strong discriminants for identifying T\.ZOs, though they provide important checks for verifying T\.ZO candidates.

\subsubsection{Mass Loss Rates}

The convective envelope of a T\.ZO is expected to be very similar to that of RSGs \citep{thorne:75,thorne:77,cannon:92,cannon:93}, so the mass loss rate of T\.ZOs has been traditionally assumed to be similar to that of RSGs. The first models of T\.ZOs \citep{thorne:77,biehle:91,cannon:92,cannon:93} did not take winds or mass loss directly into account. While our understanding of the winds of hot stars (e.g. OB main sequence or Wolf-Rayet stars) have strong theoretical bases, the mechanisms of cool star (including RSG) mass loss and winds are not well constrained \citep[and references within]{renzo:17}. RSGs mass loss rates are therefore estimated empirically (though see \citealt{kee:2021} for recent theoretical work). Commonly used and recent prescriptions for persistent rates include \citet{deJager:88,vanLoon:05,beasor:18,beasor:20,beasor:22,decin:24}, but recent works have argued for and against the magnitude and validity of these rates \citep{beasor:20,beasorsmith:22,massey:23}. In particular, RSG mass loss is known to be episodic \citep[and references therein]{massey:23}, thus prescriptions measured from instantaneous `snapshots' of RSG mass loss do not encompass the complete picture of RSG mass loss.

\citet{farmer:23} apply the \citet{vanLoon:05} mass loss rate prescription and find rates of $\dot{M}\approx10^{-5}$--$10^{-4}$ \msol\ yr$^{-1}$. This aligns with the higher end of observed RSG mass loss rates of $\dot{M}\approx10^{-7}$--$10^{-4}$ \msol\ yr$^{-1}$ \citep{mauron:11,renzo:17}. As well, \citet{farmer:23} find that should a T\.ZO undergo RSG-like hydrodynamic pulsations,  those pulsations grow in velocity and can become supersonic. This behavior was analyzed by \citet{yoon:10}, who found that these pulsations can drive `superwinds' ($\dot{M}\approx10^{-3}$--$10^{-2}$\msol yr$^{-1}$) during the late stages of RSG evolution. \citet{farmer:23} speculates that the same pulsation-driven superwinds could occur for T\.ZOs. While these superwinds may lead to extreme dust-enshrouding as a potential T\.ZO observable, they would also greatly decrease the expected lifetime of T\.ZOs.

\subsubsection{Variability}\label{variability}

The photometric variability of T\.ZOs had not been well examined from a theory perspective until recently, though \citet{thorne:75} speculated that the pulsations common in the envelopes of RSGs could also occur in T\.ZOs (e.g. pulsation driven mass loss). Many T\.ZO candidates (\S\ref{ch7:candidates}) display photometric variability.

\begin{figure}[htbp]
  \centering
  \includegraphics[width=0.75\textwidth]{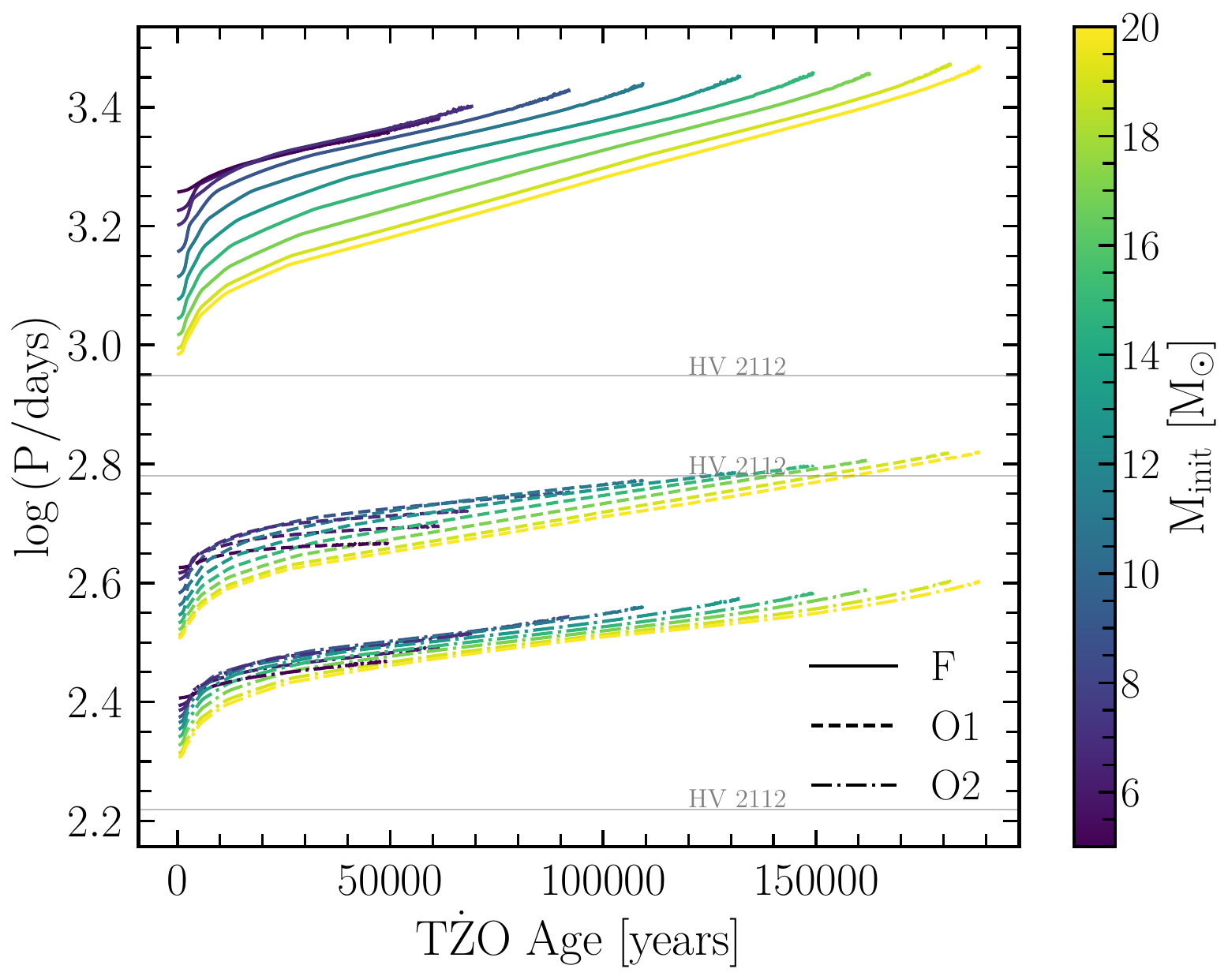}
  \caption{Period-age diagram showing the pulsation property evolution over time of the \citet{farmer:23} models. The fundamental, first overtone, and second overtone are solid, dashed, and dot-dashed lines, respectively. Initial mass is indicated by color. OGLE periods of the T\.ZO candidate HV 2112 are indicated with grey lines, but see also \S\ref{sec:hv2112}.
    This figure is from \cite{farmer:23}, with permission.}
  \label{fig:farmer23pulsations}
\end{figure}

\citet{farmer:23} used GYRE \citep{GYRE} to simulate hydrostatic pulsations for their models, and find extremely long ($\geq$1000 days) fundamental periods and first/second overtones that are $\sim$half/a quarter as long, respectively. These periods increase with time, and higher mass models have shorter periods than lower mass T\.ZOs. These are shown in Figure \ref{fig:farmer23pulsations}. The period of the fundamental mode is most sensitive to the initial helium fraction. Therefore, under these model assumptions, the period could be a unique way to constrain the He composition of future T\.ZO candidates. The \citet{farmer:23} models also lie in a relatively unique patch of the Petersen diagram ($P_{\mathrm{fundamental}}$ vs  the ratio of the first or second overtone with the fundamental). This could provide a new metric to identify T\.ZO candidates from time-domain surveys, assuming a sufficiently long baseline to detect the long fundamental periods.

\subsection{Mass}

The mass of T\.ZOs has been one of their most uncertain properties since their theoretical inception. The mass (or luminosity) gap identified by \citet{thorne:77} splits the class into `giant' and `supergiant' types (\S \ref{sec:nuclear_challenges}), thus establishing a minimum required mass for the larger supergiant T\.ZOs. \citet{thorne:77} predicted this $M_{\mathrm{min}}$ = 11.3\ \msol, while \citet{cannon:93} finds a more massive limit of $M_{\mathrm{min}}$ = 15\ \msol\ (where specifically $M_{\mathrm{env}}$=14\ \msol and $M_{\mathrm{NS}}\approx$1\ \msol when the mixing length parameter $\alpha$=1.5. The minimum mass is especially sensitive to $\alpha$, and the minimum mass can drop to as low as $M_{\mathrm{min}}\approx$10\ \msol. If the mass gap and associated minimum mass for supergiant types is real, though it's exact value is uncertain, it provides a critical check for any T\.ZO candidate.

Contrastingly, the models of \citet{farmer:23} do not contain a mass gap (\S\ref{sec:nuclear_challenges}), and range in mass from 5-20 \msol. Under the \citet{farmer:23} model assumptions, mass is not as constraining a property anymore. Therefore, this suggests that mass-based predictions are not robust against systematic uncertainties due to the modeling choices.

Constraining the current mass of stars is challenging, but attempts to estimate the mass of candidate T\.ZOs (\S\ref{ch7:candidates}) through modeling the spectroscopy \citep{levesque:14}, the spectral energy distribution \citep{beasor:18:tzo}, the current pulsation mass \citep{ogrady:20}, and the luminosity with cluster membership arguments \citep{tabernero:21} have been made.

\subsection{Astrometric and Environmental Observables}

\subsubsection{Kinematics}

Regardless of model, T\.ZO formation requires a supernova explosion to create the NS. For all channels, except on the dynamical formation scenario in a dense stellar cluster, the tight binary must remain bound after the explosion. Any subsequent system velocity therefore stems from recoil from mass loss during the supernova. Some authors find that the impact of the NS kick specifically is small compared to the total mass of the system \citep{liu:15}, while others find that the systemic velocity of the system does scale with the NS kick \citep{renzo:19}. In the HMXB common envelope evolution channel, \citet{vandenHeuvel:00} found recoil velocities range from $\approx$10--80 km s$^{-1}$, increasing for higher mass systems. If instead the T\.ZO  is formed through the NS receiving a natal kick that sends it into its companion, \citet{1994ApJ...423L..19L} find an expected velocity of $\approx$75 km s$^{-1}$.

\citet{NathanielTZOs} used a rapid population synthesis code to simulate population level demographics for T\.ZOs. This included how the NS natal kick affects the velocity of the subsequent T\.ZO system and how it might be distinguished from non-T\.ZO stars that were unbound from their binary systems due to kicks. \citet{NathanielTZOs} find velocities ranging from $\approx$10-100 km s$^{-1}$, and find that in most cases the distribution of T\.ZO system velocities is degenerate with that of unbound stars, but when the donor star in the mass transfer phase is a post-MS or Helium MS star, the velocity-mass distribution is somewhat unique from that of unbound stars.

\citet{ogrady:23} examined these predictions in the context of the Magellanic Clouds, within which several searches for T\.ZOs have occurred. Velocities of $\approx$80 km s$^{-1}$ correspond to proper motions of 0.27 and 0.34 mas yr$^{-1}$ in the SMC and LMC, respectively. At the distance of the Clouds, deviations in proper motion compared to the bulk rotation and motion of the Clouds are detectable down to $\lessapprox$0.1 mas yr$^{-1}$, so T\.ZO kicks may be detectable in high velocity cases. Low velocity recoil scenarios will remain difficult to identify, though future releases of \emph{Gaia} will significantly improve proper motion uncertainties, which will increase the range of detectable velocities.

\subsubsection{Local Star Formation Histories}

Analyses of the local star formation histories (SFHs) around astronomical objects have been used to constrain the ages stellar populations and supernova remnants \citep{badenes:10,zapartas:17,williams:18,auchettl:19,sarbadhicary:21}. In scenarios where strong constraints on local SFHs exist (e.g. in the Magellanic Clouds \citealt{HZ04,HZ09}) one can estimate the age of the stars in the vicinity of a T\.ZO candidate. Given a relatively short lifetime of the T\.ZO phase of $\approx$10$^{4}$--10$^{6}$ years \citep{cannon:93,biehle:94,farmer:23}, the total lifetime of the system should be dominated by the lifetime of the primary star. This star must be massive enough to explode and form a NS, thus we generally expect T\.ZOs to exist in areas of relatively recent star formation, though we note that local SFHs and populations must be modeled with binary stars in mind \citep[e.g.,][]{BPASS1,BPASS2}

\subsubsection{Supernova Remnants}

The supernova that produces the NS component of the eventual T\.ZO would also leave behind a supernova remnant (SNR). Typical SNR fade times are $\approx6\times10^{4}$ yr. This is short compared to the typical estimated lifetimes of T\.ZOs ($\approx10^{5}$-$10^{6}$ yr), but the SNR would still be detectable for young T\.ZOs \emph{if} the T\.ZO is formed via the NS being kicked into its RSG companion. In the case of a HMXB leading to a common envelope phase and the inspiral of the NS, the delay between NS formation and T\.ZO formation would be longer than this SNR lifetime.

The \emph{aftermath} of T\.ZOs systems could also produce SNRs if the T\.ZO explodes (see \S\ref{ch5:finalfates} for more detail). Here, the presence of a NS with a very long spin period within a SNR could indicate the descendant of a T\.ZO \citep{liu:15}.

\subsection{Spectroscopy}\label{sec:spectro}

The presence of a NS core leads to complex nucleosynthesis not seen inside of typical RSGs (\S\ref{sec:nuclear_challenges}). Material burned in the hot atmosphere of the NS is convected to the surface through the cooler envelope of the T\.ZO. Significant work has gone into determining what elements \emph{unique} to T\.ZOs might be observable through spectroscopy.

\begin{figure}[htbp]
  \centering
  \includegraphics[width=0.495\textwidth]{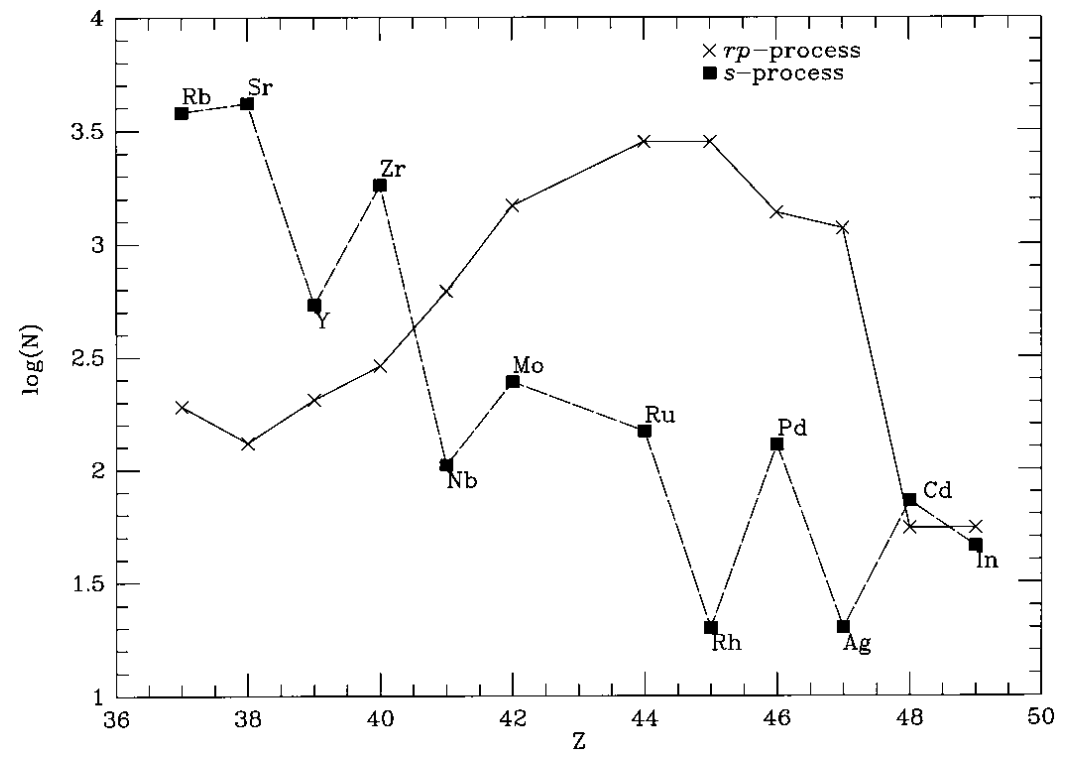}
  \vspace*{2pt}\includegraphics[width=0.495\textwidth]{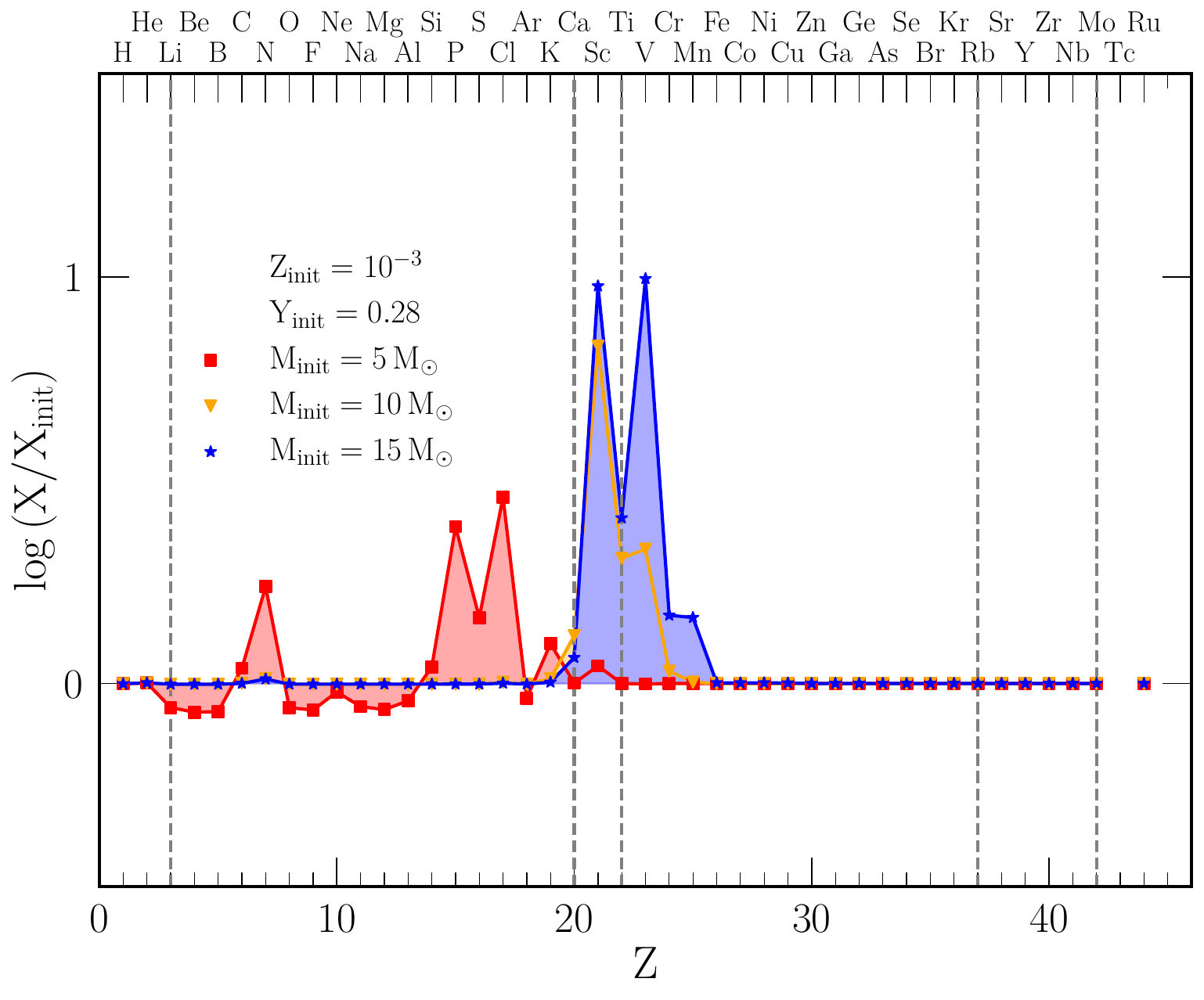}
  \caption{
  \textit{Left:} Expected abundances for irp-process (crosses, expected for T\.ZOs, from \citealt{biehle:94}) and s-process (squares, expected for red giants, from \citealt{1987ApJ...321..832M}) This panel is from \citet{vanture:99}.
  \textit{Right:} The surface composition at 10,000 years post-formation relative to initial composition for the \citet{farmer:23} models, at Z = 10$^{-3}$ and M$_{\mathrm{initial}}$ = 5\msol (red), 10\msol (orange), and 15\msol (blue). Compared to the left panel, no enhancement in the surface abundances of heavy metals (e.g. Rb, Mo) is expected. This panel is from from \citet{farmer:23}, with permission.
  }
  \label{fig:spec_on}
\end{figure}

\citet{thorne:77} first speculated that T\.ZOs should have abnormal surface abundances of certain elemental isotopes, relative to otherwise identical RSGs, but the details were not fully ironed out until future models. The concept was further developed with the inclusion of the irp process \citep{biehle:91,cannon:92,cannon:93} and eventually \citet{biehle:94} presented a full list of likely observable elements (see also Figure \ref{fig:spec_on}, left panel). We note, however, that all of these models do not account for rotation. Elements with major enhancements expected at the surface compared to the Sun are Br, Rb, Y, Nb, and especially Mo. These enhancements build up with age.  \citet{biehle:94} state that a spectral resolution of $\sim$15,000R should be sufficient to distinguish T\.ZO abundances from those of regular RSGs. \citet{podsiadlowski:95} also demonstrated that the internal environment of T\.ZOs are well suited to the production of $^{7}$Li via the Cameron-Fowler mechanism \citep{cameron:71}. A strong Li enhancement should be visible in T\.ZOs quite rapidly after formation, reach a maximum after $\approx10^{5}$ yr, and remain high throughout the T\.ZO lifetime. Finally, \citet{vanparadijs:95} suggests that detecting T\.ZO abundances anomalies in mm wavelengths would be easier than in optical wavelengths. The SiO molecule in particular is expected to be enhanced in T\.ZOs.

Attempts to identify stars with abnormally high surface abundances of the elements identified by \citet{biehle:94} and \citet{podsiadlowski:95} formed the basis for many future candidate searches (see \S\ref{ch7:candidates} for candidate-specific details). Spectroscopy of cool evolved stars, whether they are giants, supergiants, or T\.ZOs, is challenging, however. Obtaining absolute abundance measurements is nearly impossible due to the forests of metal lines and strong TiO banding.

As well, Li and heavy metal enhancement are not indicators for T\.ZOs alone. Massive (M$\geq$4\ \msol) AGB stars (including `super'-AGB stars with M=6.5-12\ \msol \citealt{garciaberro:94}) also show enhancements in Li and many of the other elements through s-process nucleosynthesis \citep{karakas:14,doherty:14,doherty:14ii,doherty:17,karakas:18,ritter:18}, though subtle differences in expectations between T\.ZOs and massive AGB stars may exist \citep{tout:14,ogrady:23}. Li enhancement peaks early then fades sharply in massive AGB stars, so concurrent strong heavy metal and Li enhancement is unexpected for massive AGBs but expected in T\.ZOs. As well, Mo is not expected to be as strongly enhanced as lower atomic number elements, such as Rb, in massive AGBs, whereas for T\.ZOs Mo production dominates over other heavy elements.

Finally, we review the new models of \citet{farmer:23}. Final nucleosynthetic yields change dramatically with initial metallicity (Z$_{\mathrm{init}}$) and He fraction (Y$_{\mathrm{init}}$) due to the availability of seed nuclei for the irp-process stemming from changes to the convective `knee' temperature. Enhancement patterns similar to that of \citet{cannon:93,biehle:94} can occur but is highly dependent on Y$_{\mathrm{init}}$ in particular. Mo is not a good element for T\.ZO detection in almost any case. Most critically, for Milky Way, LMC, and SMC metallicities, \emph{little to no metal enhancement exists} (see Figure \ref{fig:spec_on}, right panel). This brings into question previous candidates identified in the Local Group based on their apparent abundance enhancements. However, \citet{farmer:23} suggests an alternative spectroscopic signature to search for T\.ZOs -- $^{44}$Ti, specifically in the TiO and TiO$_{2}$ molecules that are prominent in the atmospheres of cool stars. See \S\ref{sec:nuclear_challenges} for more details. \citet{farmer:23} note, however, that improved molecular line lists of $^{44}$TiO and $^{44}$TiO$_{2}$ are required before detection of an enhanced $^{44}$Ti/$^{48}$Ti ratio is an effective detection method for T\.ZOs.

\subsection{Gravitational Wave Signals}

Several authors have speculated on the possible detection of gravitational wave signatures stemming from T\.ZOs, both their formation and their existence.

\citet{nazin:95} first speculated on what gravitational radiation might occur from the formation of a T\.ZO. They predict strains of h$\approx$10$^{-23.5}$ at frequencies of 10$^{-5}$--10$^{-1}$ Hz. \citet{demarchi:21} point out, however, that the frequencies are too low to detect with advanced-LIGO, and the strain is not strong enough to detect with LISA. However, \citet{moranfraile:23} speculate that T\.ZO formations may be detectable with future decihertz observatories such as DECIGO or BBO.

\citet{demarchi:21} instead focus on the possibility of observing the continuous signal from the NS moving through spacetime as it descends towards the center of the RSG, its braking inside the atmosphere leading to gravitational wave emission through energy loss. They find that the asymmetries and rapid rotation of the NS core lead to a spindown strain that could be detected by advanced-LIGO. They suggest that gravitational wave searches for T\.ZOs should focus on RSG-rich clusters within the Milky Way, such as RSGCs 1-5 and Alicante 10.

Finally, \citet{2021ApJ...919..128R} found that certain CEE scenarios, including those that may lead to T\.ZOs, may be detectable with LISA

\section{Candidate Thorne-\.Zytkow Objects}\label{ch7:candidates}

The theory of T\.ZOs can only be confirmed if their existence is proven. Searches for T\.ZOs have long attempted to do this, using observational predictions (\S\ref{ch6:observables}) to identify and analyze strong candidates. Observational prospects to find T\.ZOs are presently exciting, as new and upcoming wide-field time domain (e.g. ZTF, ASAS-SN, TESS, Rubin LSST), astrometric (\textit{Gaia}), and multi-wavelength/spectroscopic (Swift, Chandra, HST, JWST, UVEX, and various upcoming ELTs) surveys are offering us a never-before-seen view of resolved stellar populations in both our own galaxy and nearby galactic neighbors. These new advances in observational astronomy has emboldened searches for T\.ZOs. In this section we will review candidate T\.ZOs, pre-T\.ZO systems, and post-T\.ZO systems from the literature.

\subsection{U Aquarii \& VZ Sagittarii}

The first spectroscopic search for T\.ZOs came from \citet{vanture:99}, a quarter-century after T\.ZOs were first theorized. \citet{vanture:99} examined the stars U Aquarii (U Aqr) and VZ Sagittarii (VZ Sgr), both previously identified as R Coronae Borealis (RCB) stars. RCB stars are through to be produced by the merger of two white dwarfs, but \citet{iden:96} also suggested that a white dwarf-NS merger could also form an RCB star (thus this was \emph{not} a search for the `traditional' RSG+NS merger product). U Aqu had been identified as having unusually strong light s-process element signatures \citep{bond:79}. \citet{vanture:99} took higher resolution spectroscopy of U Aqu and another RCB star with a similarly peculiar spectrum, VZ Sgr, to search for evidence of rp-process and Li enhancement. They find that neither star shows the enhancement of rp-process elements expected of T\.ZOs, instead finding agreement with s-process expectations. While this search did not result in T\.ZO candidates, it helped set the standard for spectroscopic techniques, particularly examining line ratios, for future works.

\subsection{IO Per \& BD+55 388}

\citet{kuchner:02} obtained spectroscopy of 59 galactic red supergiants to search for the elemental enhancements expected of T\.ZOs. Based on previous work predicting that 1-10\% of RSGs could be T\.ZOs \citep{podsiadlowski:95}, they predict that a few T\.ZOs should be within the sample. They measure the `pseudo-equivalent widths' of absorption lines of interest -- in order to avoid the difficulty of identifying the `true' continuum in the RSG spectra, which are dominated by strong line blanketing and broad TiO bands -- and took the ratio of lines expected to be enhanced (in this case, Rb) compared to those that should not be enhanced (such as Ni). Two stars from the sample displayed enhanced Rb/Ni ratios compared to the rest of the RSGs -- IO Per and BD+55 358 (or V595 Cas). However, as mentioned in \S\ref{sec:spectro} and as the authors note, Rb can also be formed by the s-process, thus this analysis is not enough to conclusively state that IO Per and BD+55 358 are strong T\.ZO candidates. Further spectroscopic analysis to look for evidence of heavier rp-process elements (such as Mo) and Li enhancement would be required.

\subsection{HV 2112}\label{sec:hv2112}

To date, HV 2112, a cool and luminous star in the Small Magellanic Cloud (SMC), has been the most well studied T\.ZO candidate.

\subsubsection{Initial identification}

\begin{figure}[htbp]
  \centering
  \includegraphics[width=0.49\textwidth]{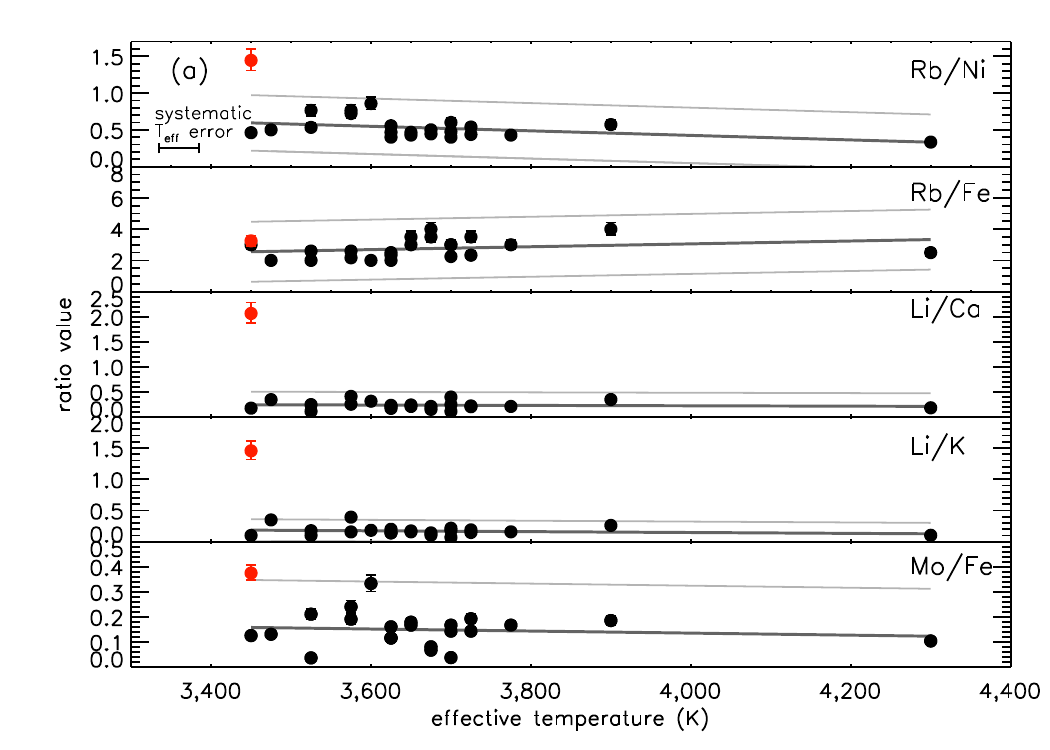}
  \includegraphics[width=0.49\textwidth]{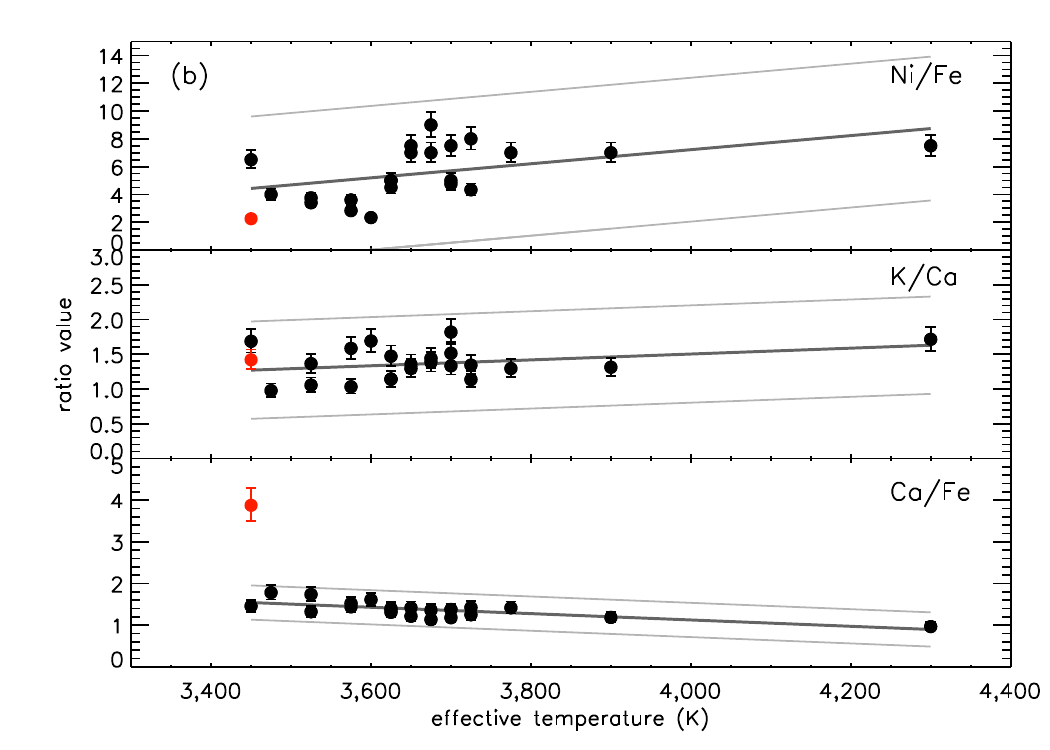}
  \caption{Effective temperature versus pseudo-equivalent width line ratios for the SMC targets of \citet{levesque:14}. Ratios with elements expected to be enhanced in T\.ZOs are in the left (a) panel, ratios with control elements are in the right (b) panel. Dark and light grey lines show the linear best fit and 3-$\sigma$ deviations to the RSG sample (black points). HV 2112 is the red point.
    This figure is from \citet{levesque:14}, with permission.}
  \label{fig:levesque2014elements}
\end{figure}

\begin{figure}[htbp]
  \centering
  \includegraphics[width=0.75\textwidth]{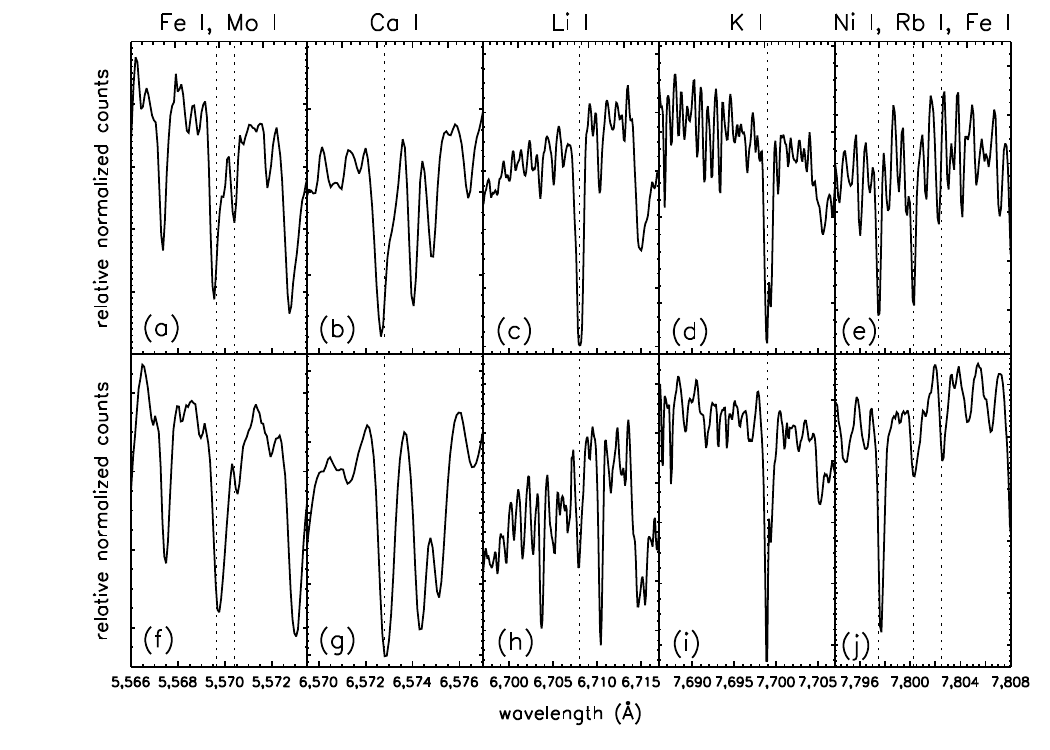}
  \caption{Spectroscopy of HV2112 (top panels) and an SMC RSG (M2002 SMC 005092, bottom panels). Lines that should be enhanced in T\.ZOs (Mo I, Li I, and Rb I) as well as control lines (Fe I, Ca I, K I, and Ni I) are highlighted.
    This figure is from \citet{levesque:14}, with permission.}
  \label{fig:levesque2014elements2}
\end{figure}

It was first identified as a T\.ZO candidate by \citet{levesque:14}. They obtained high resolution spectroscopy of several RSGs -- 24 in the Milky Way, 16 in the Large Magellanic Cloud (LMC), and 22 in the SMC -- and searched for evidence of T\.ZO-like abundance enhancements in Rb, Mo, and Li using the same `pseudo-equivalent width' line ratio technique as \citet{kuchner:02}. They also calculate line ratios comprised solely of elements not expected to be enhanced in T\.ZOs (e.g. Ni/Fe, K/Ca) as controls. See Figures \ref{fig:levesque2014elements}-\ref{fig:levesque2014elements2}. One star in their sample stood out -- HV 2112. At the time, no other star had ever been observed with both an apparent enhancement in Rb, Mo, and Li. Further, a \emph{lack} of enhancement in a Ba II line indicated that the s-process was not behind the strange abundance pattern of HV 2112. \citet{levesque:14} also identified some spectral oddities that could not be explained by a T\.ZO origin -- a discrepancy in the apparent Rb enhancement from the Rb/Ni and Rb/Fe line ratios, and an apparent enhancement in Ca/Fe, though Ca is not expected to be enhanced in T\.ZOs (but see \citealt{ogrady:23}, below). Using MARCS (Model Atmospheres with a Radiative and Convective Scheme) stellar atmosphere models, \citet{levesque:14} derive a luminosity of M$_{\rm{bol}}$ = -7.8 $\pm$0.2 (5.02), T$_{\rm{eff}}$ = 3450 K, spectral type M3I, log g = 0.0, a strong mass loss rate from high A$_{V}$ values and an excess of UV flux in the spectrum, and an initial mass of $\approx$15\ \msol, all consistent with T\.ZO predictions at the time.

\subsubsection{SMC star or galactic interloper?}

Questions have been raised about whether HV 2112 is truly located in the SMC, or whether it is a foreground (and thus much dimmer) dwarf star. \citet{levesque:14} measured a radial velocity (RV) using the Ca II triplet of $\approx$157 km s$^{-1}$ for HV 2112, consistent with it being in the SMC. However, \citet{maccarone:16}, using data from the Southern Proper Motion catalog, inferred a proper motion of over 3000 km s$^{-1}$ if it was actually located in the SMC. They suggest that HV 2112 is instead a galactic S star, and its odd spectral features could be explained by a former AGB companion polluting its atmosphere.

However, several other works have argued that HV 2112 is a true SMC member. \citet{worley:16} analyzed the contradictory proper motion estimates at the time and resolved that all properties of HV 2112 were consistent with it being an SMC member. Later, \citet{mcmillan:18} and \citet{ogrady:20,ogrady:23} independently used data from the astrometric satellite \emph{Gaia}'s data releases to show that the proper motion of HV 2112 was consistent with it being a member of the SMC.

\subsubsection{Further studies - T\.ZO or AGB star?}

Following the identification of HV 2112 as the first strong T\.ZO candidate by \citet{levesque:14}, several studies further analyzed the properties of HV 2112 to ascertain its true identity. This had led to several conflicting results.

\citet{tout:14} examined whether HV 2112 was a T\.ZO or a super-AGB (sAGB) star. sAGB stars straddle the transition between low-mass (producing planetary nebulae and CO white dwarfs) and high-mass (producing core-collapse supernovae and NSs or black holes) stellar evolution. They are the highest mass stars (6.5$\lessapprox$M$_{\odot}\lessapprox$12) that do not explode as core-collapse supernovae; at the low mass end they produce O-Ne-Mg white dwarfs, and at very thin sliver of the high mass end, they are the progenitors of electron-capture supernovae \citep[and references therein]{miyaji:80,poelarends:08,poelarends:17,doherty:17}. There are many observational similarities predicted between sAGB stars and T\.ZOs, including their otherwise odd surface abundances. The s-process in sAGB stars can create similar abundance patterns of heavy elements and lithium predicted for T\.ZOs \citep{lau:11,karakas:14,tout:14}, but see also \citet{ogrady:23} and below for some potential subtle differences. sAGBs can also reach the luminosity of HV 2112 quoted by \citet{levesque:14} \citep{smartt:02,seiss:10}. \citet{tout:14} honed in on the strange Ca abundance detected by \citet{levesque:14} as the possible discerning factor. Neither the irp- or s-process is predicted to synthesize Ca, but \citet{tout:14} point out that the Ca could have been synthesized during the formation phase of the T\.ZO, when the core of the RSG merges with the NS. Therefore, \citet{tout:14} suggests that a T\.ZO identity is more likely for HV 2112.

\citet{beasor:18:tzo} re-examined HV 2112 with further archival data and concluded \emph{against} a T\.ZO identity. They found a slightly lower maximum luminosity of log(L/\lsol) = 4.91 from a blue to mid-IR spectral energy distribution fit and accounting for the variability of the star. As well, when directly comparing the spectrum of HV 2112 to stars with very similar luminosities and spectral classes,  \citet{beasor:18:tzo} found no evidence of heavy element or Ca enrichment, though they do find a Li enrichment (see \citealt{ogrady:23} however, suggesting these comparison stars may be massive or superAGB stars and thus have similar nucleosynthesis signatures as T\.ZOs). When comparing to models from the binary population synthesis code BPASS, they suggest the luminosity and temperature of HV 2112 could instead be explained by HV 2112 being a thermally pulsing massive ($\approx$5\ \msol) AGB star.

\begin{figure}[htbp]
  \centering
  \includegraphics[width=0.95\textwidth]{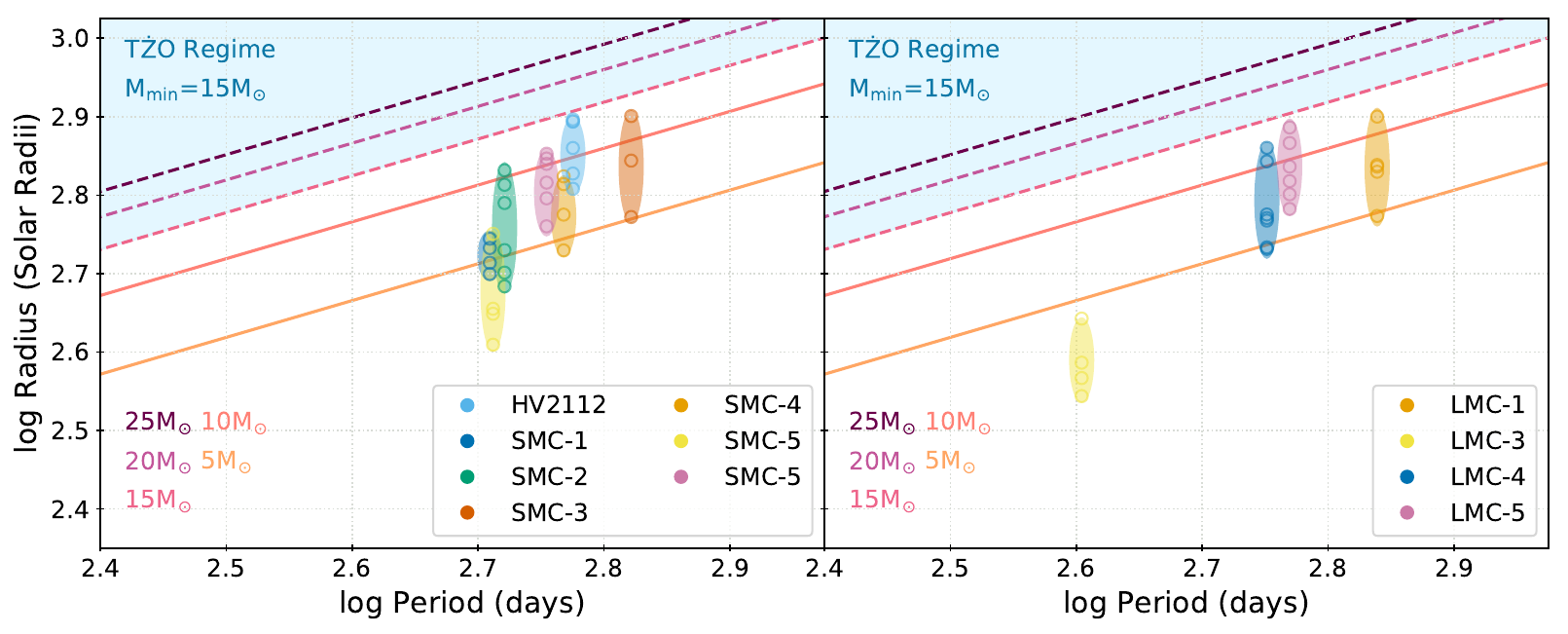}
  \caption{Period vs radius diagrams with lines of constant mass from 5 to 25\ \msol in increments of 5\ \msol, for the SMC (left) and LMC (right). HV 2112 is the light blue oval in the left plot. The other stars are likely massive or super-AGB stars. The area above 15\ \msol, the M$_{\mathrm{min}}$ from \citet{cannon:92,cannon:93} is shaded blue.
    This figure is from \citet{ogrady:20}, with permission.}
  \label{fig:ogrady2020mass}
\end{figure}

\citet{ogrady:20,ogrady:23} investigated whether HV 2112 is a unique star within the Magellanic group, or if it is part of a larger population of similar stars. They focused on the strong variability of HV 2112 -- with $\Delta$m$_{\rm{V}}$=4.8 and a $\approx$600 day period \citep{paynegaposchkin:66,soszynski:04,soszynski:11,jayasinghe:18,jayasinghe:20} -- as an important property, and identified 11 other stars in the Clouds with the same luminosity, color, and pulsational properties as HV 2112 \citep{ogrady:20}. After conducting a series of photometric, kinematic, environmental, and spectroscopic analyses of the stars, \citet{ogrady:20,ogrady:23} find: i) the 11 `HV 2112-like' stars all appeared to be a part of the same class of objects, and ii) their properties are most consistent with a sAGB star identity, but iii) while HV 2112 has similar properties, it remained an exceptional outlier in most analyses and too ambiguous to strictly classify. Specifically, the current pulsational mass of HV 2112 appears to be $\approx$10-13\ \msol \citep{ogrady:20} (see Figure \ref{fig:ogrady2020mass}), which would be inconsistent with a T\.ZO identity assuming the $M_{\mathrm{min}}$ = 15 limit of \citet{cannon:93} but could be consistent if that limit were lower. Alternatively \citet{ogrady:23} suggest the properties of HV 2112, particularly its spectrum, could also be explained by a maximally massive super-AGB star (or `hyper'-AGB star \citealt{doherty:15}).

Finally, while \citet{farmer:23} did not re-analyze HV 2112, they do comment on its consistency with their new T\.ZO models (see Figures \ref{fig:farmer23hrd} and \ref{fig:farmer23pulsations}). They find that the photometric properties of HV 2112 can be fit with a 5-8\ \msol model, but the variability of HV 2112 is difficult to fit to their pulsation models unless HV 2112 has an unobserved fundamental mode pulsation at a 1500-3000 day period. Critically, however, at SMC metallicities the \citet{farmer:23} models predict \emph{no heavy metal or Li enrichment}. Thus, if the \citet{farmer:23} models of T\.ZOs hold, the spectroscopic findings of \emph{all} of \citet{levesque:14,beasor:18,ogrady:23} would indicate that HV 2112 is not a T\.ZO.

\subsection{HV 11417}

In their paper re-analyzing HV 2112, \citet{beasor:18:tzo} also identified a new T\.ZO candidate: HV 11417, also a star in the SMC. With a high luminosity (log(L/\lsol) = 4.92) and extremely strong Rb abundance (even compared to  HV 2112 and other similar stars). They do not detect a Li enhancement, but note that strong TiO banding may have obscured the Li absorption line.

However, \citet{ogrady:20} showed that the \emph{Gaia} DR2 proper motions of HV 11417 were inconsistent with the motions of other SMC stars, indicating HV 11417 might be a foreground dwarf star. As well, under the \citet{farmer:23} models (Figure \ref{fig:farmer23hrd}), in addition to the Rb enhancement now being inconsistent with a T\.ZO identity, the luminosity of HV 11417 is too low to be matched by any model at SMC metallicity (though a 5\msol model at high Z may fit), while the variability period of HV 11417 implies a mass much too high to match its luminosity.

\subsection{VX Sgr}

VX Sgr is a cool, variable, and extremely luminous Galactic star. Previous authors classified it as either a massive/super-AGB star or a RSG \citep[and references therein]{tabernero:21}. \citet{tabernero:21} examined VX Sgr in detail in order to determine if it belonged to either of those classes, or if it was a T\.ZO. Most critically, its spectrum revealed an extremely strong Rb enhancement but no Li, which is more consistent with massive/super-AGB predictions than those of \citet{cannon:92,cannon:93} for T\.ZOs. The more recent T\.ZO mass and pulsation models of \citet{farmer:23} could match VX Sgr quite well (Figure \ref{fig:farmer23hrd}), but it's strong Rb enhancement is at odd with the nucleosynthesis predictions in these models.

\subsection{Candidate pre- and post-T\.ZO systems}

Here we note some examples of possible pre- and post-T\.ZO candidate systems identified in the literature.

\citet{manikantan:24} analyzed the orbital evolution of the high mass X-ray binary GX 301-2 and found it to have the highest fastest orbital decay known of any HMXB (P$_{\mathrm{orb}}$/\.P$_{\mathrm{orb}}\approx$0.6$\times10^{5}$ yr). They suggest it may be a future T\.ZO candidate via the common envelope path (but see \S\ref{subsec:binary_coalescence}).

The first search for post-T\.ZO objects was performed by \citet{coe:98}, who used deep infrared and optical imaging of three proposed T\.ZO remnants, specifically pulsating X-ray sources with 5-10s pulsation periods and no evidence of binarity or a detectable optical counterpart. They find that none of the objects are likely post-T\.ZO systems.

\citet{liu:15}, however, identified the X-ray source 1E161348-5055 within the young supernova remnant RCW 103 to be a potential post-T\.ZO system. It's extremely slow rotation period (6.67 hours) and slow proper motion within the SNR agree with predictions for post-T\.ZO NSs.

\section{Summary and Future Outlook}\label{ch8:summary}

In the nearly 5 decades since Thorne-\.Zytkow Objects were first theorized, our understanding of their theoretical nature, and that of the evolution of massive stars in general, has changed dramatically. Our observational capabilities have similarly expanded substantially in that time. This had led to considerable progress as a community in understanding and attempting to identify T\.ZOs, though many questions still remain. 

\textit{Formation:} It remains unclear whether T\.ZOs can assemble and form the structure originally envisioned by \citet{thorne:75,thorne:77}. The most promising formation scenario involves binary evolution,  where a star and a NS merge as a result of mass transfer or a collision. However, the extent to which the envelope is removed during the merger, and therefore the structure of the merger remnant, is still uncertain. Understanding the assembly of T\.ZOs is crucial for providing observational constraints on them. Simulating the merger between a star and a NS can offer insights into the actual structure of T\.ZOs.

\textit{Structure, Evolution, and Modeling:} From the modeling side, the complex nuclear burning and sensitivity to uncertain ingredients (e.g., mass loss) represent a numerical challenge. For this reason \tzo~models are few and far between ($\gtrsim$20\,years between published models), and no assessment of the systematic theoretical uncertainties on predicted observables exists. Nevertheless, given the subtleties of the predicted signatures and the large potential for ``false positive'' identifications, being able to test the robustness of the predictions would be a numerically challenging but welcome development. We also note that issues in the modeling of mass loss, pulsations, convection, stability, and evolution are not unique to T\.ZOs -- these are challenging problems for all cool and luminous stars, and thus the solutions to these problems will impact huge swaths of stellar astrophysics. 

\textit{Final Fates:} The possible final fates of T\.ZOs depend heavily on the modeling predictions detailed above. Depending on the existence or lack thereof of a mass gap, a T\.ZO may either collapse, resulting in an explosion observed as a long timescale SN or a pulsar, or leave behind a slow-rotating NS. 

\textit{Observables and Candidates:}  Under the model assumptions of \citet{thorne:75,thorne:77,cannon:92,cannon:93}, the most identifying characteristics of T\.ZOs are their spectroscopic signatures and mass. T\.ZOs should possess elements produced by the irp-process (e.g. Rb, Mo) and an abundance of Li which sets it apart from almost every other class of red, luminous star. Possible confusion with the most massive 'low-mass' stars (super-AGB) stars can be disentangled through estimations of the current mass, assuming the mass gap exists. The \citet{farmer:23} models instead predict no mass gap and little to no heavy metal enrichment at metallicities found within our galaxy and nearby neighbors, though $^{44}$TiO and $^{44}$TiO$_{2}$ may be promising alternatives. Several candidate T\.ZOs have been explored in the literature, though none without scrutiny. HV 2112, undoubtedly a strange star in any case, is the most well-studied and promising candidate to date.

\begin{ack}[Acknowledgments]

The authors thank Emily Levesque for her advice, discussions, and for providing excellent comments on this chapter. AJGO thanks Maria Drout and Katie Breivik for helpful discussions and advice. MR is grateful to Rob Farmer for years of stimulating interactions on stellar evolution and help with MESA, and for involving him in the first T\.ZO calculations with this code. AVG thanks Hans-Thomas Janka for useful discussions on stellar collapse and Enrico Ramirez-Ruiz for useful discussion on T\.ZOs.
\end{ack}

\seealso{\citet{thorne:75,thorne:77,cannon:92,cannon:93,biehle:94,podsiadlowski:95,levesque:14,beasor:18:tzo,ogrady:20,ogrady:23,farmer:23,2023arXiv231106741H,2024ApJ...971..132E}}

\bibliographystyle{Harvard}
\bibliography{reference}

\end{document}